\documentclass[11pt]{article}
\usepackage[utf8]{inputenc}
\usepackage[T1]{fontenc}
\pdfoutput = 1

\usepackage{amsmath}
\usepackage{amssymb}
\usepackage{bbm}
\usepackage{bbold}
\usepackage{braket}
\usepackage{caption}
\usepackage{cite}
\usepackage{color}
    \definecolor{darkgreen}{rgb}{0,0.5,0}
    \definecolor{darkred}{rgb}{0.5,0,0}
    \definecolor{darkblue}{rgb}{0,0,0.6}
    \definecolor{purple}{rgb}{0.4,.2,0.7}
\usepackage[mathcal]{eucal}
\usepackage{float}
\usepackage{graphicx}
\usepackage[hyperfootnotes = true, colorlinks = true, linkcolor = darkblue, citecolor = purple]{hyperref}
\usepackage{mathabx}
\usepackage{mathtools}
\usepackage{pdfsync}
\usepackage{slashed}
\usepackage[normalem]{ulem}
\usepackage{upgreek}
\usepackage{url}

\usepackage[margin = 2.2cm]{geometry}
\usepackage[ragged]{footmisc}
\def\be{\begin{equation}}
\def\ee{\end{equation}}

\def\ben{\begin{equation}}
\def\een{\end{equation}}

\let\a=\alpha \let\b=\beta \let\g=\gamma \let\d=\delta \let\e=\varepsilon
   
\let\l=\lambda \let\m=\mu    \let\r=v
 \let\t=\tau

\let\w=\omega \let\G=\Gamma   \let\L=\Lambda

\def\be{\begin{equation}}
\def\ee{\end{equation}}
\def\ba{\begin{array}}
\def\ea{\end{array}}

\def\dalemb#1#2{{\vbox{\hrule height .#2pt
       \hbox{\vrule width.#2pt height#1pt \kern#1pt
               \vrule width.#2pt}
       \hrule height.#2pt}}}

\newcommand{\bea}{\begin{eqnarray}}
\newcommand{\eea}{\end{eqnarray}}

\let\tilde=\widetilde

\def\diag{{\rm diag}}

\def\Lc{\mathcal{L}}
\def\Wc{\mathcal{W}}
\def\PSL{\textsf{PSL}}
\def\SL{\textsf{SL}}
\def\d{{\rm d}}
\def\Pf{{\rm Pf\,}}
\def\Hit{{\rm Hit}}

\renewcommand{\d}{\mathrm{d}}
\renewcommand{\i}{\mathrm{i}}
\renewcommand{\S}{\textsf{S}_0}
\renewcommand{\tilde}{\widetilde}
\DeclareMathOperator{\Tr}{Tr}

\numberwithin{equation}{section}

\begin{document}

\thispagestyle{empty}
\begin{center}
    ~\vspace{5mm}
    
    {\Large \bf 
    
    Higher spin JT gravity and a matrix model dual}
    
    \vspace{0.4in}
    
    {\bf Jorrit Kruthoff}

    \vspace{0.4in}

   Department of Physics, Stanford University, Stanford, CA 94305-4060, USA
    
    {\tt kruthoff@stanford.edu}
\end{center}

\vspace{0.4in}

\begin{abstract}

We propose a generalization of the Saad-Shenker-Stanford duality relating matrix models and JT gravity to the case in which the bulk includes higher spin fields. Using a $\PSL(N,\mathbb{R})$ BF theory we compute the disk and generalization of the trumpet partition function in this theory. We then study higher genus corrections and show how this differs from the usual JT gravity calculations. In particular, the usual quotient by the mapping class group is not enough to ensure finite answers and so we propose to extend this group with additional elements that make the gluing integrals finite. These elements can be thought of as large higher spin diffeomorphisms. The cylinder contribution to the spectral form factor then behaves as $T^{N-1}$ at late times $T$, signaling a deviation from conventional random matrix theory. To account for this deviation, we propose that the bulk theory is dual to a matrix model consisting of $N-1$ commuting matrices associated to the $N-1$ conserved higher spin charges. 

We find further evidence for the existence of the additional mapping class group elements by interpreting the bulk gauge theory geometrically and employing the formalism developed by Gomis et al. in the nineties. This formalism introduces additional (auxiliary) boundary times so that each conserved charge generates translations in those new directions. This allows us to find an explicit description for the $\PSL(3,\mathbb{R})$ Schwarzian theory for the disk and trumpet and view the additional mapping class group elements as ordinary Dehn twists, but in higher dimensions.

\end{abstract}

\pagebreak

\tableofcontents

\section{Introduction}

In recent years an increasingly intimate connection between gravity and random matrix theory in the context of the AdS/CFT correspondence has emerged. In particular, in the case of two-dimensional JT gravity a precise and exact mapping between the two has been established \cite{Saad:2019lba}. Observables on the gravity side, which are bulk path integrals with certain boundary conditions \cite{Saad:2018bqo, Saad:2019lba, Goel:2020yxl, Stanford:2019vob, Yan:2022nod, Stanford:2021bhl, Iliesiu:2021ari}, are in one-to-one correspondence with averages of correlation functions of partition functions, computed in random matrix theory. 

The beauty of this correspondence is that the topological expansion of the gravity theory maps onto the known topological expansion in random matrix theory and thus allows for a potential non-perturbative definition of the gravity theory. For various theories related to JT, for instances ones that are supersymmetric or with more general dilaton potentials, a similar correspondence has been derived \cite{Stanford:2019vob, Witten:2020wvy, Turiaci:2020fjj, Maxfield:2020ale}, and also some proposals have been made in higher dimensions \cite{Cotler:2020ugk}, although there the topological expansion is under much less control. 

The relation between 2d gravity theories and random matrix theory also brings about its fair share of paradoxes, such as the factorization and discreteness puzzle \cite{Maldacena:2004rf, Blommaert:2021fob, Blommaert:2021gha, Saad:2021rcu, Saad:2021uzi, Blommaert:2022ucs, Johnson:2022wsr, Johnson:2021zuo}. Some of these have been resolved recently, but the UV interpretation and implementation of the proposed solutions remains elusive. This is so not only because the UV complete theory one whishes to consider is complicated and not known in its entirety, but also because the 2d theory it induces would be one with matter fields. These cause various problems on higher-genus surfaces and so makes the construction of a matrix model dual more intricate, but see upcoming work by Jafferis et al. Nevertheless, understanding this case in detail is crucial for understanding the dictionary between gravity and RMT ensembles in more detail \cite{Belin:2020hea, Belin:2021ryy, Schlenker:2022dyo, Chandra:2022bqq}. 

In this paper we want to offer an explicit duality between matrix models on one side and a bulk two dimensional gravity with fields beyond the usual dilaton and metric. The theory in the bulk we will consider contains not only the metric and dilaton, but also higher spin fields with spin up to some integer $N>2$. The metric formulation of this theory is rather complicated, but it has a simple description in terms of a $\PSL(N,\mathbb{R})$ BF theory. This theory can be studied rather explicitly and we will show how to compute the partition function on simple topologies. Some of these calculations have already been done in the past \cite{Gonzalez:2018enk, Narayan:2019ove, Datta:2021efl} in 2d, but we will focus on formulating the calculation in any Riemann surface, which will not only require a definition of the trumpet but also an understanding of the underlying moduli space integrals. One of the main objectives will be to calculate the leading connected gravitational contribution to spectral form factor. 

Another motivation for this work is that in the past we have seen that higher spin theories play an intriguing role in holography and it is worthwhile to understand what the status in two dimensions is, as one might hope to have a lot of analytic control over in particular the topological expansion. The study of higher spin gravity theories and its relation to holography has a long and exciting history. This started with the work of Vasiliev \cite{VASILIEV1990378, Vasiliev:2003ev} and was implemented into holography by Klebanov and Polyakov in \cite{Klebanov:2002ja} in AdS$_4$ by relating it to an $O(N)$ vector model. See also \cite{Anninos:2011ui} for a realisation in four dimensional de Sitter space. These theories all contained an infinite number of higher spin fields as required by internal consistency. The higher spin theories were also further explored in three dimensions \cite{Gaberdiel:2010pz} by constructing duals of minimal model CFTs or by using a Chern-Simons formulation \cite{Gaberdiel:2011wb, henneaux2010nonlinear, Campoleoni:2010zq, Gutperle:2011kf}. There one can also study a finite number of spins, which will be more closely related to what were are about to discuss, but we will digress on the infinite number of spins case as well towards the end.

\subsection{Summary of results}

The structure and results of the paper are as follows. In section \ref{sec:setup} we will start with setting up the problem and introduce the relevant notation and boundary conditions. We will argue that the partition functions are one-loop exact as a result of the Duistermaat-Heckman theorem. In section \ref{sec:disktrumpet} we then present the explicit expressions for the disk partition function and give a definition and expression for the trumpet partition function. In the sections thereafter we present our main findings,
\begin{enumerate}
    \item In section \ref{sec:gluing} we consider higher-genus contributions, in particular the wormhole or double trumpet geometry. For this we first determine the right moduli space and how it can be parametrized, which mimicks the conventional Fenchel-Nielsen parametrization of Teichmüller space, but with the important difference that additional coordinates beyond the Fenchel-Nielsen ones are needed to cover the entire moduli space. For the wormhole, however, such complications do not arise and one needs to integrate over $N-1$ length and twist variables. The integrals over the twist variables diverges and in analogy with the resolution to this in the JT case, where one quotients by the mapping class group, we propose a similar quotient that makes the integrals finite. We then calculate the spectral form factor and its leading order connected component. As a function of time it behaves as sketched in Fig. \ref{fig:SFF}. In particular at late times we get a change in the ramp behaviour, 
    \be 
    Z_{\rm Cyl}(\b,T) \sim T^{N-1}\, , 
    \ee
    which stands in sharp contrast with conventional random matrix theory predictions. 
    
    \begin{figure}
        \centering
        \includegraphics[scale=0.5]{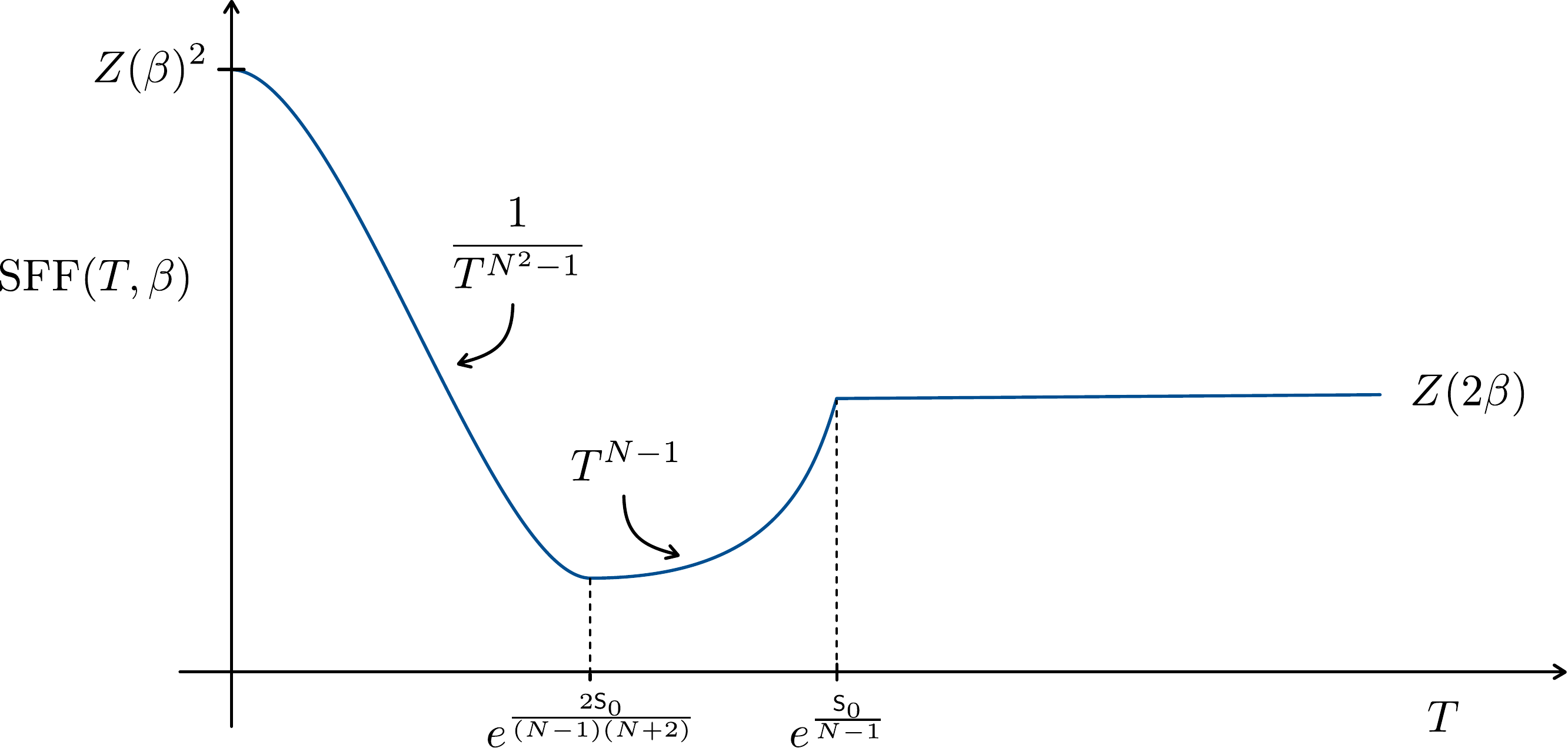}
        \caption{Sketch of the spectral form factor (SFF) as a function of time $T$ at fixed temeprature $\beta$ in the $\PSL(N,\mathbb{R})$ BF theory. At early times the disconnected contribution dominates and gives rise to a $T^{1-N^2}$ power law decay, followed by a $T^{N-1}$ power law growth after the dip time which is approximately $e^{2\S/(N+2)(N-1)}$. The signal then plateaus around the Heisenberg time, which is $e^{\S/(N-1)}$ and follows from the proposed matrix model dual.}
        \label{fig:SFF}
    \end{figure}
    
    \item We show that this behaviour for the cylinder can be naturally explained by a matrix model containing $N-1$ commuting matrices, one matrix for each conserved charge on the boundary. This model is equivalent to a Dyson gas of $L$ particles not in one dimension but in $N-1$ dimensions and with a repulsive force given by the logarithm of the $N-1$ dimensional distance between two particles. 
    
    \item Section \ref{sec:PSL3Schw} is devoted to finding a suitable geometric description, following \cite{Gomis:1994rz}, of the higher spin theories that in previous sections was formulated purely in terms of gauge theory variables. Along the way we will find the suitable (non-linear) version of the Schwarzian actions for the $\PSL(3,\mathbb{R})$ disk and trumpet. Furthermore, with this geometric description we give more evidence for the existence of higher-spin generalizations of the Dehn twists.
    
    \item To obtain more evidence for our duality between the higher spin theory and a matrix model of commuting matrices we would also need to consider higher genus corrections. These corrections are much more complicated to calculate and from the bulk no known recipe, such as Mirzakhani's recursion relations is known. Our matrix model could be a first step in that direction and we offer some thoughts and future directions on that in section \ref{sec:discussion}. We also comment there on the large $N$ limit of our model, its relation to SYK and present some speculative ideas for the generalisation of the double cone geometry in Lorentzian signature.

\end{enumerate}

\section{$\PSL(3,\mathbb{R})$ BF theory}\label{sec:setup}

In this section we will present the main example we will be working with; an $\PSL(3,\mathbb{R})$ BF theory. We will first study the classical theory and discuss the boundary conditions and then discuss the exact evaluation of the partition function. Even though we work here with a $\PSL(3,\mathbb{R})$ theory, most of the statements generalize to $\PSL(N,\mathbb{R})$. See also \cite{Campoleoni:2010zq, Campoleoni:2011hg, Narayan:2019ove, Saad:2019lba, Gonzalez:2018enk}. 

The action of BF theory on a two dimensional manifold $\Sigma$ is given by 
\be \label{S1}
I = -\S \chi(\Sigma) - \i \int_{\Sigma} \Tr BF.
\ee
Here the first term is a topological term we have put as a genus counting parameter so that path integrals on different Riemann surfaces are weighted by a factor $e^{\S \chi(\Sigma)}$.\footnote{One might be tempted to write this topological term in terms of gauge theory variables as well. In the $N=2$ case this can be done, since as we will see later, the bundles one considers have a first Chern class equal to the Euler character. For $N>2$ this doesn't hold anymore because the bundles do not have integer valued invariants. Furthermore, since we are interested in surfaces with boundary one would need to generalize to that case as well. For these reasons we simply put the Euler character as in \eqref{S1} and refrain from writing it purely in terms of the gauge theory variables.} $B$ is a scalar transforming in the adjoint of $\PSL(3,\mathbb{R})$, so if $g \in \PSL(3,\mathbb{R})$, then
\be 
B \mapsto g^{-1} B g\, .
\ee
Furthermore, $F$ is the curvature two-form of a $\mathfrak{sl}(3,\mathbb{R})$ gauge field $A$,
\be 
F = \d A + A \wedge A.
\ee
The gauge transformations act on $A$ as 
\be 
A \mapsto g^{-1}A g + g^{-1} \d g 
\ee
Since $B$ acts as a Lagrange multiplier in the action \eqref{S1}, the gauge field $A$ must be flat, i.e. $F=0$. The other equation of motion follows from varying with respect to $A$ and gives
\be 
0=DB = \d B - [B,A]
\ee
In what follows it will be important to consider $\Sigma$ to have a boundary in which case we need to supplement \eqref{S1} with a boundary term. In particular we will focus on disk and cylinder topologies. In the $\PSL(2,\mathbb{R})$ case, i.e. usual JT gravity, the boundary condition is chosen so that it agrees with Dirichlet boundary conditions in the gravitational variables. This boundary conditions gives the Schwarzian as the boundary theory and consequently the disk (and trumpet) partition function are one-loop exact. Specifically the Euclidean boundary action generates `time' translations (in the symplectic sense \cite{Stanford:2017thb}) and we want to generalize this to the $\PSL(3,\mathbb{R})$ (and $\PSL(N,\mathbb{R})$). This can be achieved by simple keeping the same boundary condition and boundary term as in the $\PSL(2,\mathbb{R})$ case, \cite{Blommaert:2018oro}. This amounts to modifying \eqref{S1} to
\be \label{action2d}
I = -\S \chi(\Sigma) - \i \int_{\Sigma} \Tr B F + \frac{\i}{2} \int_{\partial \Sigma} \Tr B A
\ee
with boundary condition
\be \label{canonicalBC}
\left.(B + 2\i \g A)\right|_{\partial \Sigma} = 0.
\ee
where we parametrize the boundary circle with $u\sim u + \beta$ and $\g$ is a constant. With this boundary condition alone we will get a particle on the $\PSL(3,\mathbb{R})$ group manifold and our path integral will be one over $\text{Loop}(\PSL(3,\mathbb{R}))/\PSL(3,\mathbb{R})$, which is not what we want. We want to generalize the Schwarzian case, which was an path integral over $\textsf{Diff}(S^1)/\PSL(2,\mathbb{R})$. From a bulk point of view this arose from the fact that we integrate over gauge transformations that leave a certain asymptotic form of the gauge field invariant. The boundary action will then be such that it generates time translations on the boundary of $\Sigma$, which is, as mentioned above, what wish to achieve. These boundary conditions have been worked out in the context of $\textsf{SL}(N,\mathbb{R}) \times \textsf{SL}(N,\mathbb{R})$ Chern-Simons theory in 3d \cite{Campoleoni:2010zq, Campoleoni:2011hg, Gutperle:2011kf, Ammon:2011nk, deBoer:2014fra} and also more recently in 2d \cite{Gonzalez:2018enk}. For the $N=3$ case they read, 
\be \label{AAdS2bc}
A = L_0 \d r + e^{-r L_0}A_u e^{r L_0} \; \d u,\quad A_u = L_1 + L_{-1} \Lc(u) + W_{-2}\Wc(u)\quad ,
\ee
with $L_i$ and $W_i$ the generators of the $\mathfrak{sl}(3,\mathbb{R})$ which have the following commutation relations
\begin{align}
[L_i, L_j] &= (i-j)L_{i+j},\quad [L_i,W_n] = (2i-n)W_{n+i}\\
[W_m, W_n] &= - \frac{1}{3}(m-n)(2m^2+2n^2 - mn - 8)L_{m+n}
\end{align}
with $i=-1,0,1$ and $n=-2,-1,0,1,2$. Explicit matrix representations are given in appendix \ref{app:generators}. In \eqref{AAdS2bc}, $r$ is the bulk radial coordinate and $\mathcal{L}$ and $\mathcal{W}$ two arbitary smooth periodic functions. We picked the principal embedding as this corresponds to having fields with spin $2$ to $N$ in the bulk \cite{deBoer:2014fra, Castro:2012bc}.\footnote{From the representation theory of $\PSL(N,\mathbb{R})$ this means that the adjoint representation of $\PSL(N,\mathbb{R})$ decomposes into $N-1$ $\PSL(2,\mathbb{R})$ representations of dimension $(2j+1)$ with $j=1,\dots N-1$.} We will also see that when considering higher genus contributions, the principal embedding is the natural embedding to consider as it picks out a component of the moduli space that contain Teichmuller space.

The form \eqref{AAdS2bc} restricts the form of the allowed gauge transformation and gives rise to the boundary gravition and boundary spin-$3$ field. To see how this works we write a infinitesimal $\PSL(3,\mathbb{R})$ gauge transformation $\eta$ as
\be\label{modes1}
\eta(u) =  \sum_{i=-1}^{+1} L_i \xi_i(u) + \sum_{n=-2}^{+2} W_n \chi_i(u).
\ee
In order for the form of \eqref{AAdS2bc} to be maintained to first order in $\xi$ and $\chi$, we find that the functions $\xi_i$ and $\chi_i$ can be parametrized by two periodic functions $\xi_1(u) = \e(u)$ and $\chi_2 = \zeta(u)$,
\begin{align}\label{modes2}
\xi_0 &= -\e',\quad \xi_{-1} = \Lc \e - 8 \Wc \zeta + \frac{\e''}{2},
\quad \chi_1 = - \zeta',\quad \chi_0 = 2 \Lc \zeta + \frac{\zeta''}{2},\\\label{modes3} \chi_{-1} &= - \frac{5}{3} \Lc \zeta' - \frac{2}{3}\Lc' \zeta - \frac{\zeta'''}{6},\quad 
\chi_{-2} = \Wc \e + \left(\Lc^2 + \frac{\Lc''}{6}\right) \zeta + \frac{7}{12}\Lc'\zeta' + \frac{2}{3}\Lc \zeta'' + \frac{\zeta''''}{24}
\end{align}
and consequently the above gauge transformation changes $\Lc$ and $\Wc$ to first order as
\begin{align}
\Lc &\to \Lc + \Lc' \e + 2\Lc \e' + \frac{1}{2}\e''' - 8 \Wc' \zeta - 12 \Wc \zeta',\\
\Wc &\to \Wc + \Wc' \e + 3 \Wc \e' + \frac{8}{3} \Lc^2 \zeta' + \frac{3}{4} \Lc'' \zeta' + \frac{5}{4} \Lc' \zeta'' + \frac{8}{3} \Lc \Lc'\zeta + \frac{1}{6} \Lc''' \zeta + \frac{5}{6} \Lc \zeta''' +  \frac{1}{24}\zeta'''''
\end{align}
With the boundary conditions \eqref{AAdS2bc} we see that there are two physical modes that we need to path integrate over: $\e$ and $\zeta$. Under the gauge transformations with $\zeta = 0$, $\Lc$ transforms as a spin-$2$ field and $\Wc$ as a spin-$3$ field. Notice also that we can always do a gauge transformation to set $\Lc$ and $\Wc$ to be constant. The asymptotic symmetry algebra that we get with these boundary conditions is one copy of the $\Wc_3$ algebra, a non-linear extension of the Virasoro algebra discovered by Zamolodchikov \cite{Zamolodchikov:1985wn}. These statements also trivially generalizes to the $\PSL(N,\mathbb{R})$ case, where one would get an $\Wc_N$ asymptotic symmetry algebra. 

The fact that $\mathcal{L}$ transforms as a spin $2$ field, means that we want the integral of $\mathcal{L}$ to be the boundary action as that would be the generator of time translations. Our boundary conditions indeed achieve this, 
\be \label{boundaryaction}
I_{\partial} = \g \int_0^{\b} \d u\; \Tr A_u^2 = - 8\g \int_0^\b \d u\; \Lc(u)
\ee
but $\mathcal{L}$ is just a function of $u$ right now and not over the modes we integrate over. These modes are such that perturbatively, when expanded around the saddle point, they become the $\e$ and $\zeta$ modes. Let us call these modes $F_1$ and $F_2$. As discussed in \cite{Gonzalez:2018enk} one can always write $\mathcal{L}$ in terms of $F_i(u)$. The path integral is thus over the $F_i$ with action given by 
\eqref{boundaryaction} with $\Lc$ in terms of $F_i$. In section \ref{sec:PSL3Schw} we show explicitly what the form of this action in terms of the $F_i$ is. Notice again that all this extends to all $\PSL(N,\mathbb{R})$. 

Let us now discuss the measure for our path integral. We already noted that after we integrate $B$ out in the bulk, we get a path integral over flat $\PSL(3,\mathbb{R})$ connections. As explained in \cite{Witten:1991we, Saad:2019lba, atiyah1983yang} there is a natural measure on the space of such connections that arises from the fact that the space of flat connections can be endowed with a symplectic structure. The symplectic form is given by 
\be \label{canonicalSymplecticMeasure}
\Omega(\delta_1 A,\delta_2 A) = 2\a \int_{\Sigma} \Tr \delta_1 A \wedge \delta_2 A
\ee
with $\a$ a normalization constant and the associated volume form $\mu$ is its Pfaffian, $\mu = \Pf \Omega$. However, as we will see below, generically this symplectic form is degenerate and hence not a good symplectic form. Just as in the $\PSL(2,\mathbb{R})$ case, the associated zero modes in $F_1$ and $F_2$ can be easily projected out by quotienting the integration space by the group $G_0$ the zero modes generate. The number of zero modes depends on the geometry one considers as we will see below.  

The reason why the present discussion has been rather abstract and the main reason we went through this is because it is clear from the fact that the integration space is symplectic and the (boundary) action is the generator of `time' translations that we can apply the Duistermaat-Heckman theorem here (modulo the issue with the zero modes that can be dealt with and we will do so explicitly below). This means that the path integral over $F_i$,
\be \label{Z}
Z = \int \frac{d\mu[F_1,F_2]}{G_0} e^{-S_{\partial}[F_1,F_2]}
\ee
is one-loop exact and we can simply stick to the linearized level we started this section off with.

In the next section we will carry this calculation out in detail by studying the boundary action to second order in $\e$ and $\zeta$. We will also see that there is one additional piece of data needed to compute the partition function.   

\section{Disk and trumpet}\label{sec:disktrumpet}

This additional piece of data involves some global properties of the gauge field $A_u$, in particular its holonomy around the boundary circle. For the disk, the holonomy has to be trivial (valued in the center of $\SL(N,\mathbb{R})$), but for a (generalization of) the trumpet or defect geometries this need not to be the case.  

Thus for the disk partition function we integrate over all flat connections that preserve \eqref{AAdS2bc} asymptotically and have trivial holonomy around the Euclidean circle. We can always view such connections $A_u$ as a gauge transformation of the constant connection $A_0$, 
\be \label{Aconstant}
A_u = g(u)^{-1}A_0 g(u) + g(u)^{-1}\partial_u g(u)
\ee
with $A_0 = L_1 + \Lc L_{-1} + \Wc W_{-2}$, and $\Lc, \Wc$ constants and $g(u) \in \PSL(3,\mathbb{R})$.  The trivial holonomy condition is now satisfied when $A_0$ satisfies it. For the trumpet we can do the same, but the holonomy of $A_0$ needs to be non-trivial. 

Before we get to this however, it will be helpful to first review how it works in the $\PSL(2,\mathbb{R})$ case. In this case there is only $\Lc$ and for the disk topology we need,
\be 
\mathcal{P} \exp \oint_\mathcal{C} A = - \mathbf{1}
\ee
with $\mathcal{C}$ along the $u$ direction and the relevant component $A_u$ of $A$ is given by
\be 
A_u = \begin{pmatrix}
0 & -\mathcal{L}(u) \\
1 & 0
\end{pmatrix}.
\ee
For constant $\Lc(u)$ it easy to calculate this, because we can ignore the path ordering and we get the condition
\be 
\Lc = \frac{\pi^2 n^2}{\b^2}\, ,
\ee
with $n$ an odd integer. We pick $n=1$ as other non-zero integers correspond to higher and/or reverse windings of the boundary.\footnote{This constraint seems natural from the theory in metric variables, but is rather unnatural from the gauge theory perspective and says that a subsector of the gauge theory describes gravity. In ordinary JT gravity this is also the case and one only wants to integrate over the coadjoint orbit ${\rm Diff}(S^1)/{\textsf{SL}}(2,\mathbb{R})$. This is also related as to why the density of states is a hyperbolic sine instead of the Placherel measure of the universal cover of $\textsf{SL}(2,\mathbb{R})$, \cite{Jafferis:2019wkd, Iliesiu:2019xuh}. Another, perhaps more important reason, why to exclude other $n$ is that some quadratic fluctuations will be unstable.} For the trumpet geometry, one requires the holonomy to lie in the hyperbolic conjugacy class of $\PSL(2,\mathbb{R})$. This conjugacy class is parametrized by a single real constant $b$ and is interpreted geometrically as the length of the small end of the trumpet. One finds $\Lc = -b^2/4\b^2$. 

\subsection*{Disk}

For $\PSL(3,\mathbb{R})$ we want to pick a similar condition on the holonomy. In particular, for the disk, we simply take the holonomy around the non-contractible cycle $\mathcal{C}$ to be the identity. \footnote{In general, for $\PSL(N,\mathbb{R})$ we take the holonomy equal to $(-1)^N \mathbb{1}$.} To calculate the holonomy for $\mathcal{C}$ a non-contractible cycle of the disk, we first diagonalize $A_0$ in \eqref{Aconstant}. We find that it is given by 
\be 
A_0 = V \diag(\l_1, \l_2 , -\l_1 - \l_2)V^{\dagger}
\ee
with $\l_i$ a solution to 
\be 
-8\Wc + 4\Lc \l_i + \l_i^3 = 0,\quad \text{ for } i = 1,2\, ,
\ee
and $V \in \PSL(3,\mathbb{R})$. By exponentiating, we get the eigenvalues of the holonomy. It will be convenient to write $\Lc$ and $\Wc$ in terms of the eigenvalues of the $\l_i$,
\be \label{LWeigs}
\Lc = -\frac{1}{4}(\l_1^2 + \l_2^2 + \l_1 \l_2),\quad \Wc = - \frac{1}{8}\l_1 \l_2(\l_1 + \l_2)\, . 
\ee
From this it is easy to see that if we pick $\l_1 = - \l_2 = 2i\pi n/\b$ for $n$ integer, we get the required holonomy. For the same reason as before we set $n=1$, so $\mathcal{L} = \pi^2/\b^2$ and $\mathcal{W} = 0$.  

\subsection*{Trumpet}

For the trumpet we again pick the hyperbolic conjugacy class, but now of $\PSL(3,\mathbb{R})$ (or $\PSL(N,\mathbb{R})$). This means we have $\PSL(3,\mathbb{R})$ elements whose eigenvalues are all real. For convenience and in order to follow the literature \cite{Goldman1990, WPN3} \footnote{Here we use $\ell_i$ and in \cite{Goldman1990, WPN3} these are $\ell$ and $2m$ and $\ell$ and $m$, respectively.}, it will be convenient to parametrize them as
\be \label{defTrumpet}
\l_1 = \frac{3\ell_1-2\ell_2}{12\b},\quad \l_2 = \frac{\ell_2}{3\b}\, ,
\ee
and so 
\be 
\Lc = - \frac{1}{16\b^2}\left( \frac{\ell_1^2}{4} + \frac{\ell_2^2}{3} \right) ,\quad \Wc = -\frac{\ell_2}{384\b^3}\left(\ell_1^2 - \frac{4\ell_2^2}{9}\right)
\ee
with $\ell_1$ and $\ell_2$ real constants that generalise the length parameter $b$ in the spin two case. It is important to also specify the ranges of $\ell_1$ and $\ell_2$ as well. For $N=2$ the range of $b$ was the positive reals, because the element of $\PSL(2,\mathbb{R})$ with positive and negative $b$ are conjugate in $\PSL(2,\mathbb{R})$. For $\PSL(3,\mathbb{R})$ one can work out the ranges as well, by noting that one can permute the eigenvalues of the holonomy by conjugating with elements of Weyl group $S_3$ of $\PSL(3,\mathbb{R})$. In doing so we can always arrange $\lambda_i$ such that
$\l_1$ is the largest eigenvalue and $\l_2$ the second largest and $-\l_1 - \l_2$ the smallest. This results in $\ell_1 > 0$ and $|\ell_2| < \ell_1/2$, so the space of $\PSL(3,\mathbb{R})$ trumpets is parametrized by 
\be \label{spaceR}
\mathcal{R} = \{ (\ell_1,\ell_2)\, | \,  \ell_1 > 0 \, , \, -\ell_1/2 < \ell_2 < \ell_1/2 \} \subset \mathbb{R}^2.
\ee
For general $N$ the procedure is analogous, but one would use elements of $S_N \subset \PSL(N,\mathbb{R})$ to order the eigenvalues. 

\subsection*{Defects}

Finally, there is another conjugacy class of $\PSL(3,\mathbb{R})$ one can consider, namely one with two complex conjugate eigenvalues and one real eigenvalue. In the notation above we would then have $\l_1 = (a + \i b)/\b = \l_2^*$. These \emph{geometries} would be the generalisation of the defect geometries one encounters in the elliptic conjugacy class of $\PSL(2,\mathbb{R})$. We will not discuss this cases further in this paper.

\subsection{Troubles with geometry}\label{sec:geom1}

The gauge field is now fixed and we can study the path integrals. However, before doing that it will be instructive to say a few words about the second order formulation of the theory, even though we will not be using it until section \ref{sec:PSL3Schw}. This will perhaps give the geometry aficionados a firmer grip on the higher spin theories.

For such an interpretation we need to extract a zweibein and spin connection from the connection $A$. To do so, let us define \cite{witten19882+, Fredenhagen:2014oua, Campoleoni:2010zq}
\be 
e = \frac{1}{2}(A + A^{\dagger}),\quad \w = \frac{1}{2}(A - A^{\dagger}),
\ee
where $e$ and $\w$ contain not only the zweibein and spin connection but also the higher spin generalisations, this is so because we have $8$ generators now instead of $3$. To be more concrete, we have $3$ generators constructed out of the $L_i$ (see Appendix \ref{app:generators}),
\be 
J_0 = \frac{1}{2}(L_1 + L_{-1}),\quad J_1 = \frac{1}{2}(L_1 - L_{-1}),\quad J_2 = L_0 
\ee
and five generators that we write as
\begin{align}
J_{11} &= \frac{1}{4}(W_2 + W_{-2} - 2 W_0),\quad J_{22} = W_0\\
J_{01} &= \frac{1}{4}(W_2 - W_{-2}),\quad J_{02} = \frac{1}{4}(W_1 + W_{-1}),\quad J_{12} = \frac{1}{4}(W_1 - W_{-1})\,.
\end{align}
The tensor $J_{\a\b}$ is symmetric and traceless (raised and lowered with $\eta_{\a\b}$). The reason we have chosen these particular generators is because of hermiticity. In the spin-$3$ case we have three independent components of $\w$. This is so because we have a two-dimensional local Lorentz index $a=1,2$ (raised and lowered with $\delta_{ab}$) and so the spin$-2$ spin connection component of $\w$, which we write as $\w^{ab}_\mu$, has one independent component $\w_0$ since due to antisymmetry in the Lorentz index. The spin$-3$ component has three indices and is denoted by $\w^{abc}_\mu$ and has symmetries $\w^{abc} = -\w^{cba} = \w^{bac}$. Hence there are two independent components which we denote by $\w^{01}$ and $\w^{02}$. So we have three components to $\w$, meaning that we need $3$ antihermitian generators of $\mathfrak{sl}(3,\mathbb{R})$, which are the $J_0$, $J_{01}$ and $J_{02}$ generators. The remaining five are hermitian and couple to the einbein $e^a_\mu$ and the spin$-3$ generalisation $e^{ab}_\mu$ (which is symmetric in its indices). The gauge field is thus expanded as
\be 
A = e^a J_a + \w^0 J_0 + e^{ab}J_{ab} + \w^{0a}J_{0a}
\ee
and the vanishing of the curvature gives rise to $8$ conditions on these components, generalizing the usual torsion constraint and the expression for the Ricci curvature. The metric and spin$-3$ field $\phi$ in the bulk are given by traces of $e$, 
\be \label{metricSpin3}
g_{\mu\nu} = \frac{1}{2}\Tr(e_\mu e_\nu),\quad \phi_{\mu\nu\rho} = \frac{1}{3!} \Tr (e_\mu e_\nu e_\rho)
\ee
We now know what the relation is between the gauge theory variables and the metric variables, but such a relation is incomplete without knowing what gauge transformations correspond to what transformations in the metric variables. In fact, the whole difficulty of trying to interpret higher spin theories geometrically is because it is in general hard to make such an identification. Some gauge transformations, however are easily identified as diffeomorphisms. For instance, consider the infinitesimal gauge variation with parameter $\eta$ of $e$ and $\w$,
\be 
\delta_{\eta} e = \d \eta_+ + [e,\eta_-] + [\w,\eta_+],\quad \delta_{\eta}\w = \d \eta_- + [\w,\eta_-] + [e,\eta_+]\, ,
\ee
with $\eta_{\pm} = 1/2(\eta \pm \eta^{\dagger})$. From these variations it is clear that a diffeomorphism generated by the vector field $\xi^{\mu}$ corresponds to those $\eta$ that can be written as (suppressing the frame index)
\be 
\eta_+ = \xi^{\mu} e_\mu,\quad \eta_- = 0.
\ee
Trying to do the same for the spin$-3$ transformation (which we know at the linear level from \eqref{metricSpin3}) is much more complicated. 

A perhaps more fruitful discussion is to analyse the symmetries. For instance, let us focus on the disk. As we discussed in the previous sections, the disk is given by the gauge field configuration,
\be 
A = J_2 \d r + \left(e^r - e^{-r}\frac{\pi^2}{\b^2}\right)J_1 \d u + \left(e^r + e^{-r}\frac{\pi^2}{\b^2}\right)J_0 \d u
\ee
This is a pure gauge configuration $g^{-1} \d g$ with $g = e^{uL_1 + u \pi^2/\b^2 L_{-1}}e^{rL_0}$. One can easily check that variations of this gauge field with the parameter,
\be 
\eta_A = (g^{-1}T_A g)^B T_B
\ee
leaves the connection invariant. Here $A,B$ is a collective index for the $8$ generators. These correspond to the $3$ Killing vectors and $5$ Killing tensors of AdS$_2$, but to find them explicitly, one would need to invert the object $\eta_A = S^{M}_A \xi_M$ with $M = \{\mu,(\nu\rho)\}$, but $S^{M}_A$ is not a square matrix. See \cite{Fredenhagen:2014oua} for a proposal for how to proceed. 

For the trumpet one can do exactly the same, but now one has $\mathcal{W} \neq 0$. The metric and spin$-3$ field take the form,
\be 
\d s^2 = \d r^2 + \left( (e^r - \mathcal{L} e^{-r})^2 + 4 e^{-4r}\mathcal{W}^2 \right)\d u^2,\quad \phi = \mathcal{W}e^{-2r}(e^r - \mathcal{L}e^{-r})^2 \d u^3
\ee
with $\mathcal{L}$ and $\mathcal{W}$ given by their trumpet values. The gauge connection for this case is invariant under two gauge transformations, since they now also need to commute with the non-trivial holonomy of $A$. One of these two symmetries is time translations $\partial_u$ (accompanied by a local Lorentz transformation) and corresponds to the gauge transformation \eqref{modes1} with $\e(u)$ set to a constant and $\zeta = 0$. The other symmetry corresponds to setting $\zeta$ to a constant and $\e$ to zero. This is some combination of the Killing vectors $\partial_u$ and the Killing tensors $g^{uu}\partial_{u}^2$ and $\partial_u^2$ and involves the spin$-3$ field $\phi$ for the trumpet. We will find a more appealing geometric interpretation of this second symmetry in section \ref{sec:PSL3Schw}. 

\subsection{Saddles and fluctutations}

So far we have discussed what the constraints on the connection $A_u$ are and what their holonomy is in case of the disk and trumpet geometry. To carry out the exact path integral evaluation, we need to understand the saddle points of the action \eqref{boundaryaction} and the fluctuations around them. 

The saddle points are not too complicated to figure out. Looking at the remaining equations of motion to be solved, we have 
\be 
\partial_r B - [B,A_r] = 0,\qquad \partial_u B - [B,A_u] = 0
\ee
The first equation is easily solved by plugging in \eqref{AAdS2bc} and yields $B = e^{-r L_0}B_0(u) e^{r L_0}$. Inserting this in the second equation, using \eqref{AAdS2bc}, and noticing that the $r$ dependence drops out so that we can evaluate it on the boundary $r \to \infty$ where we have \eqref{canonicalBC}, we get
\be 
\partial_u A_u = 0 \Rightarrow \mathcal{L}' = 0 = \mathcal{W}'
\ee
Another way to see this is to vary the action with respect to the modes we found in the previous section. This gives,
\be 
\delta I = -8\int_0^\b du \left(\e(u)\Lc'(u) + 2 \Lc(u)\e'(u) + \frac{1}{2}\e'''(u) - 8 \zeta(u)\Wc'(u)-12 \Wc(u) \zeta'(u) \right)
\ee
with $\e, \zeta$ periodic with period $\b$. This needs to vanish for arbitrary (periodic) $\e$ and $\zeta$, so we see that $\Lc' = \Wc' = 0$ on the saddle point. Thus the saddle points are connections with $A_u$ constant. This means that on the saddle points we simply have to deal with the connections $A_0$ discussed around \eqref{Aconstant} and study flucutations around them. This argument trivially extends to $\PSL(N,\mathbb{R})$. 

The fluctuations are easily computed by considering the modes in \eqref{modes1}, \eqref{modes2} and \eqref{modes3} and expanding the action $I_{\partial}$ to second order in $\e$ and $\zeta$, we get the following action for $\e$ and $\zeta$,
\be \label{S2}
I_{(2)} = \g \int_0^\b \d u \left[ 2(\e'')^2 - 8 \Lc (\e')^2 + 96 \Wc \e' \zeta' + \frac{2}{3}( (\zeta''')^2 - 20 \Lc (\zeta'')^2 + 64 \Lc^2 (\zeta')^2) \right]. 
\ee
Notice that when $\zeta$ and $\Wc$ are set to zero, we recover the usual linearisation of the Schwarzian action. In the case of the disk, we have $\Wc=0$ and $\Lc = \pi^2/\b^2$ and the action has the following zero modes,
\be 
\e(u) = 1,\; e^{\pm 2\pi \i / \b},\quad \zeta(u) = 1,\; e^{\pm 2\pi \i / \b},\;  e^{\pm 4\pi \i / \b}.
\ee
coming from the $\PSL(3,\mathbb{R})$ symmetries discussed above. For the trumpet we have zero modes only for constant $\zeta$ and $\e$. 

Since our partition function is one-loop exact, this action for the quadratic fluctionation is enough to calculate the full partition function. To carry that computation out, we first need to calculate the symplectic measure of the fluctuations $\e$ and $\zeta$ using \eqref{canonicalSymplecticMeasure}. To do so we insert $\delta_i A = d\Theta_i + [A,\Theta_i])$ and write the symplectic measure as a boundary integral,
\be
\Omega(\delta_1 A,\delta_2 A) = 2\a \int_{\partial \Sigma} \d u \Tr \Theta_1 (\d \Theta_2 + [A,\Theta_2])
\ee
We take $\Theta_i$ to be \eqref{modes1} with \eqref{modes2} and \eqref{modes3}, but with two different $\e_i$ and $\zeta_i$. The algebra is straightforward and results in 
\begin{align}
\Omega(\delta_1 A,\delta_2 A) = 2\a \int_0^\b \d u &\left[ -8\Lc \e_1(u)\e_2'(u) + 48 \Wc (\e_1(u)\zeta_2'(u) + \zeta_1(u) \e_2'(u)) + 2 \e_1'(u)\e_2''(u)\right.\nonumber\\
&\quad \left.+ \frac{2}{3}(\zeta_1''(u)\zeta_2'''(u) - 20 \Lc \zeta_1'(u)\zeta_2''(u) + 64 \Lc^2 \zeta_1(u)\zeta_2'(u)  \right].
\end{align}
As a two-form, we can write this as \footnote{With this form of $\Omega$ we can also explicitly check  that $\iota_{V}\Omega = \d H$ with $H$ equal to $I_{(2)}$ modulo the constants $\g$ and $\a$ and $V$ the vector field that generates 'time' translations: $\delta \e = \e'$ and $\delta \zeta = \zeta'$.}
\begin{align} \label{symplecticMeasure_ezeta}
\Omega = 2\a \int_0^\b \d u &\left[ -4\Lc \d \e(u) \wedge \d \e'(u) + 24 \Wc (\d \e(u) \wedge d\zeta'(u) + \d \zeta(u) \wedge \d \e'(u)) + \d \e'(u)\wedge \d \e''(u) \right.\nonumber\\
&\quad \left.+ \frac{1}{3}(\d \zeta''(u)\wedge \d\zeta'''(u) - 20 \Lc \d \zeta'(u) \wedge \d \zeta''(u) + 64 \Lc^2 \d \zeta(u)\wedge \d \zeta'(u)  \right].
\end{align}
To find the measure for the $\e$ and $\zeta$ modes, we need to find its Pfaffian, which we do by employing the following orthonormal basis on the circle,
\be \label{modeexpansion}
\e(u) = \sum_{|n|\geq n_0} e^{-\frac{2\pi \i n u}{\b}} (\e_n^R + \i \e_n^I),\quad \zeta(u) = \sum_{|m|\geq m_0} e^{-\frac{2\pi \i m u}{\b}}(\zeta_m^R + \i \zeta_m^I)\, ,
\ee
with $n_0$ and $m_0$ are integers. For the disk we have $n_0 = 2$ and $m_0 =3$, whereas for the trumpet $n_0 = m_0 = 1$. This projects out the zero modes and makes $\Omega$ non-degenerate. We will assume that $m_0 \geq n_0$. The modes $\e(u)$ and $\zeta(u)$ are real, so $\e_n^R = \e_{-n}^R$ and $\e_n^I = - \e_{-n}^I$ and analogously for $\zeta_n^{I,R}$. Using this mode expansion, we can write the symplectic form as,
\begin{align}
\Omega &= 2\a \left( \sum_{|n|\geq n_0} \frac{16\pi^3 n }{\b^2}\left(n^2 - \frac{\b^2 \mathcal{L}}{\pi^2} \right) \d\e_n^R \wedge \d \e_n^I \right.\nonumber\\
&\left. + \sum_{|n|\geq m_0} \frac{64\pi^5 n}{3\b^4} \left(n^2 - \frac{4\Lc \b^2}{\pi^2}\right)\left(n^2 -  \frac{\Lc \b^2}{\pi^2}\right) \d \zeta_n^R \wedge \d \zeta_n^I\right.\nonumber\\
&\left. + \sum_{|n|\geq m_0} 96\pi \Wc n \left( \d \e_n^R \wedge \d \zeta_n^I - \d \e_n^I \wedge \d \zeta_n^R\right)\right)\, .
\end{align}
In terms of the modes $\e_n^{I,R}$ and $\zeta_n^{I,R}$ the symplectic form is a direct sum of four by four matrix $\Omega_{ab}$ ($a,b=1,\dots,4$ label the four modes for fixed $n$) for $n\geq m_0$, but contains two by two blocks for $n_0 \leq n < m_0$. The Pfaffian is then the product of the Pfaffian of each block. Using that for a four by four matrix, ${\rm Pf\,}M_{ab} = M_{12}M_{34} - M_{13}M_{24} + M_{14}M_{23}$, we deduce that the symplectic measure for the fluctuation modes is given by
\begin{align}
{\rm Pf\,}\Omega = \prod_{n\geq m_0} &\left( \frac{2^{12} \a^2 \pi^8}{3\b^6} \right)\left[n^2 \left(n^2 - \frac{4\Lc \b^2}{\pi^2}\right)\left(n^2 -  \frac{\Lc \b^2}{\pi^2}\right)^2 - \frac{27\Wc^2 n^2 \b^6}{\pi^6}\right]\d\e_n^R \d\e_n^I \d\zeta_n^R \d\zeta_n^I \times\nonumber\\
&\quad \times \prod_{n_0 \leq n < m_0} \frac{32\a \pi^3}{\b^2} n \left(n^2 - \frac{\b^2 \Lc}{\pi^2} \right)\d\e_n^R \d\e_n^I \d\zeta_n^R \d\zeta_n^I\, .
\end{align}
The action \eqref{S2} can also be written in terms of the modes \eqref{modeexpansion} and reads
\begin{align}
I_{(2)} &= 2\g \sum_{n\geq n_0} \frac{32 n^2 \pi^4}{\b^3} \left(n^2 -  \frac{\Lc \b^2}{\pi^2}\right)\left( (\e_n^R)^2 + (\e_n^I)^2 \right) + \nonumber\\
&\quad + \sum_{n\geq m_0} \left[ \frac{384 n^2 \pi^2 \Wc}{\b} (\e_n^R \zeta_n^R + \e_n^I \zeta_n^I) + \frac{128 \pi^2 n^2}{3\b^5} \left(n^2 - \frac{4\Lc \b^2}{\pi^2}\right)\left(n^2 -  \frac{\Lc \b^2}{\pi^2}\right)\left( (\zeta_n^R)^2 + (\zeta_n^I)^2 \right) \right]\, .
\end{align}
We now have everything in place to calculate \eqref{Z}. The gaussian integrals are straightforward to carry out\footnote{If we were to pick $\mathcal{L}$ such that some of the gaussians have the wrong sign, this integral would not converge and would be ill-defined for such $\mathcal{L}$. This happens, for instance, when we pick $\l_1 = -\l_2 = 2\pi \i n/\b$ with $n\neq 1$ below \eqref{LWeigs}.}  and (unsurprisingly) cancel against the measure factor coming from the Pfaffian up to some simple factors, 
\be 
Z(\b) = e^{8\g \b \Lc} \prod_{n\geq m_0} \frac{\b^2 \a^2}{4\g^2 n^2} \prod_{n_0 \leq n < m_0} \frac{\b \a}{2\g n} = e^{8\g \b \Lc} \frac{\Gamma(n_0)}{\Gamma(m_0)} \left( \frac{\b \a}{2\g} \right)^{m_0-n_0} \prod_{n\geq m_0} \frac{\b^2 \a^2}{4\g^2 n^2}\, .
\ee
The infinite product can be evaluated using $\zeta$-function regularisation, 
\be 
\prod_{n\geq m_0}  \frac{\b^2 \a^2}{4\g^2 n^2} = \prod_{1\leq m<m_0} \frac{4\g^2 m^2}{\b^2 \a^2} \prod_{n\geq 1}  \frac{\b^2 \a^2}{4\g^2 n^2} = \left(\frac{2\g \,\G(m_0)}{\b \a}\right)^{2(m_0-1)} \frac{\g}{\pi \b \a}\, ,
\ee
and we finally arrive at
\be 
Z(\b) = \frac{1}{2\pi} \G(n_0)\G(m_0)^{2(m_0 - 3/2)} \left( \frac{2\g}{\b \a} \right)^{m_0 + n_0 - 1} \, e^{8\g \b \Lc}\, .
\ee
For the disk we have $\Lc = \pi^2/\b^2$, $n_0 = 2$ and $m_0 = 3$ and hence the partition function becomes,
\be 
Z_{\rm Disk}(\b) = 4 e^{\S} \left( \frac{2\g}{\b \a} \right)^4 e^{\frac{8\g \pi^2}{\b}},
\ee
where we added the factor $e^{\S}$ since the disk has Euler character one. For the trumpet we have $\mathcal{L} = -\frac{1}{16\b^2}(\ell_1^2/4 + \ell_2^2/3)$ and $n_0 = 1 = m_0$,
\be 
Z_{\rm Trumpet}(\b;\ell_1,\ell_2) = \frac{\g}{\pi \b \a} e^{-\frac{\g}{2\b}\left(\ell_1^2/4 + \ell_2^2/3 \right)}\, .
\ee

\subsection{General $N$}

\begin{figure}
    \centering
    \includegraphics[scale=0.27]{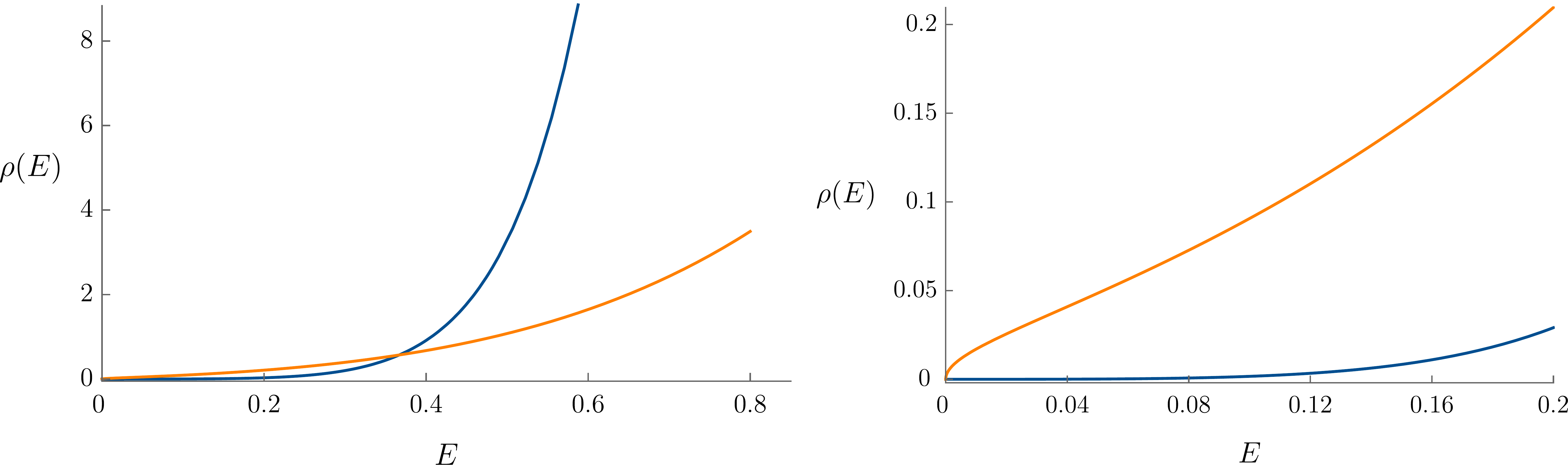}
    \caption{\textbf{Left:}Spectral density for the $N=3$ theory (blue) compared to the one for the Schwarzian theory (orange). The spectral density is larger at high energies. It grows like $e^{4\pi \sqrt{E}}$ as compared to $e^{2\pi \sqrt{E}}$ for the Schwarzian theory. \textbf{Right:} Same spectral densities at low energies to accentuate the fact that for odd $N$ the density goes to zero smoothly (in this case like $E^3$) as compared to even $N$, where it goes to zero like a root.}
    \label{fig:N3SpectralDensity}
\end{figure}

Let us also mention the results for general $N$. For the disk, motivated by the considerations in three dimensions \cite{Castro:2011iw}, we want the eigenvalues of $A_u$ to be 
\be
{\rm Eigen}(A_u) = \frac{\pi\i}{\b} (\pm 1, \pm 3, \dots, \pm (N-1)/2) 
\ee
for $N$ even and 
\be
{\rm Eigen}(A_u) = \frac{2\pi\i}{\b} (\pm 1, \pm 2, \dots, \pm (N-1)/2, 0) 
\ee
for $N$ odd. The on-shell action for the disk will then be (for both $N$ even and odd)
\be 
S_{\rm on-shell} = -\frac{\g \pi^2}{3\b}N(N^2-1)\, .
\ee
In case of the trumpet we wanted all the eigenvalues of $A_u$ to be real. Let us denote them for simplicity by $\l_i = \ell_i/2\b$ for $i = 1,\dots N-1$ and $\l_N = -\sum_{i=1}^{N-1}\ell_i/2\b$. The on-shell action is then 
\be 
S_{\rm on-shell} = \frac{\g}{2\b}\sum_{i\leq j} \ell_i\ell_j\, .
\ee
For the one-loop factor we simply note that for the disk we have $N^2-1$ zero modes and $N-1$ for the trumpet. The full partition functions are therefore given by,
\be 
Z_{\rm Disk}(\b)\,\, \propto \,\, \frac{e^{\S}}{\b^{\frac{N^2-1}{2}}} e^{\frac{\g \pi^2}{3\b}N(N^2-1)},\quad Z_{\rm Trumpet}(\b)\,\, \propto \,\, \frac{1}{\b^{\frac{N-1}{2}}} e^{-\frac{\g}{2\b}\sum_{i\leq j} \ell_i\ell_j}\, .\label{ZdiskZtrumpet}
\ee
The associated density of states of these partition functions is given by 
\be \label{DOS}
\rho_{\rm Disk}(E)\; \propto \; e^{\S}E^{\frac{N^2-3}{4}} I_{\frac{N^2-3}{2}}\left( 2\pi \sqrt{\frac{\g (N^2 - 1)E}{3}} \right),\quad \rho_{\rm Trumpet}(E)\; \propto\; \left(\frac{E}{Q}\right)^{\frac{N-3}{4}}J_{\frac{N-3}{2}}\left( \sqrt{2\g Q E} \right)
\ee
with $Q = \sum_{i\leq j} \ell_i \ell_j$. At small energies, we know from the JT gravity case that $\rho_{\rm Disk}(E)$ goes like $\sqrt{E}$, which is already an interesting hit towards a hermitian one-matrix model dual. For JT with higher spin fields, we actually have a slightly different behaviour depending on whether $N$ is even or odd. Expanding $\rho_{\rm Disk}$ in \eqref{DOS} at small $E$ we get
\be \label{DOSexpansion}
\rho_{\rm Disk}(E) \sim E^{\frac{N^2-3}{2}} + \dots\,.
\ee
For even $N$ this has the familiar root singularity, but for odd $N$ the density of states vanishes smoothly at $E=0$. See Fig. \ref{fig:N3SpectralDensity} for the spectral density for the $N=3$ case compared to the one for the Schwarzian theory.
When we discuss our proposal for the matrix model dual of JT with higher spin fields in section \ref{sec:matrixmodel}, we will see that this feature naturally comes up as well. 

Besides the two conjugacy classes we considered now, when $N$ becomes bigger many more different conjugacy classes exist. For general $N$, since $A_u$ has real entries, the eigenvalues can be real or come in complex conjugate pairs. The trumpet geometry could thus also be defined in a completely different way by, for instance, taking only two of the eigenvalues to be real. It turns out however, when we start considering higher genus contributions to the path integral, one should indeed take all eigenvalues to be real.

\section{Higher Teichmüller theory and wormholes}\label{sec:gluing}

So far we focussed on the simplest topologies and only a single asymptotic boundary. To understand the effects of higher spin fields on random matrix statistics, we need to go beyond this and compute two-boundary observables to which geometries like the double trumpet or wormhole geometry contribute. For JT gravity this is rather straightforward to do as one only needs to employ techniques from Teichmüller theory. In particular, one constructs higher genus contributions to the path integral by gluing bordered Riemann surfaces to trumpets. This gluing requires two coordinates, one for the relative angle between the boundaries that are being glued and the length of the boundary. These two coordinates are known as the Fenchel-Nielsen twist and length coordinates. The measure for this gluing originates from the Weil-Petersson symplectic form on the moduli space of bordered Riemann surfaces. In terms of the Fenchel-Nielsen coordinates, this symplectic form in the case of the moduli space of Riemann surfaces with genus $g$ and $n$ geodesic boundaries takes the simple form \cite{10.2307/2374363}
\be 
\omega_{\rm WP} = \sum_{i=1}^{3g-3+n} \d\ell_i \wedge \d\tau_i 
\ee
The twist and length coordinates are thus canonically conjugate. The non-triviality, which highlights the difference between JT gravity and $\PSL(2,\mathbb{R})$ BF theory, is the range of the twist variables $\t_i$ and the particular connected component of the space of flat $\PSL(2,\mathbb{R})$ connections one is supposed to pick. 

In the case of $\PSL(2,\mathbb{R})$ the space of flat $\PSL(2,\mathbb{R})$ connections contains $4g-3$ different components and the one relevant for JT gravity is the one for which the flat $\PSL(2,\mathbb{R})$ bundle over $\Sigma$ has Chern number $2g-2$ \cite{HITCHIN1992449, Stanford:2019vob}. This is to ensure that the connection gives rise to a hyperbolic metric on the genus $g$ surface\cite{Witten:2020bvl} . The range of the twist variables is then fixed (or compactified) as follows. In JT gravity, geometries related to each other by a large diffeomorphism, i.e. the action of the mapping class group ${\rm MCG}(\Sigma)$ of the surface $\Sigma$ under considerations, should not be double counted. In the BF theory we thus need to consider the usual moduli space of flat $\PSL(2,\mathbb{R})$ connections (modulo gauge transformations) but now quotiented by ${\rm MCG}(\Sigma)$. This results in a compact range of the twist variables $\t_i \sim \t_i + \ell_i$. For the $\PSL(N,\mathbb{R})$ theory we propose to follow the exact same logic, albeit with some differences, in part because we do not have the geometric picture as we had for $\PSL(2,\mathbb{R})$. 

The moduli space of flat $\PSL(N,\mathbb{R})$ connections on a surface $\Sigma_{g,n}$ with genus $g$ and $n$ holes is given by 
\be \label{ModuliSpace}
\mathcal{M}_{g,n}^N = {\rm Hom}(\pi_1(\Sigma_{g,n}), \PSL(N,\mathbb{R}))/\PSL(N,\mathbb{R}).
\ee
As noted by Hitchin \cite{HITCHIN1992449} this space has $3$ connected components when $N$ is odd and $6$ when $N$ is even. Picking the right component is not entirely trivial, since the condition using the Chern number of the $\PSL(2,\mathbb{R})$ bundle does not work anymore for $N>2$. This is because the maximal compact subgroup of $\PSL(N,\mathbb{R})$ is $\textsf{PSO}(N,\mathbb{R})$ and does not have integer valued topological invariants, but are rather $\mathbb{Z}_2$-valued. Instead, as discussed by Hitchin \cite{HITCHIN1992449}, one picks an embedding of $\PSL(2,\mathbb{R})$ in $\PSL(N,\mathbb{R})$ given by the (unique, up to conjugation) $N$-dimensional representation of $\PSL(2,\mathbb{R})$ in $\PSL(N,\mathbb{R})$. Using this embedding, one can construct a homomorphism $\rho_0: \pi_1(\Sigma) \to \PSL(N,\mathbb{R})$ and the component we pick is the component that contains $\rho_0$. This component is called the Hitchin component, which we will denote by $\Hit_{g,n}^N$. Notice that the Hitchin component is precisely determined by the principal embedding of $\PSL(2,\mathbb{R})$ in $\PSL(N,\mathbb{R})$.

Let us describe this component in a bit more detail. Hitchin showed that $\Hit_{g,n}^N$ is diffeomorphic to $\mathbb{R}^{(2g-2+n)(N^2-1)}$ (for $2g-2+n>0$). The dimensionality simply follows from counting the allowed holonomies modulo overall gauge transformations on a surface with $n$ holes and genus $g$. To parametrize this component, we can proceed just as in the $\PSL(2,\mathbb{R})$ theory, but instead of assigning one length coordinate to each hole, we assign $N-1$ length coordinates $\ell^{(j)}_i$ with $i = 1,\dots, N-1$ to the $j$-th hole. These are the generalisations of the Fenchel-Nielsen length-coordinates. In contrast to $\PSL(2,\mathbb{R})$ theory, they are not enough to parametrize the full moduli space. There are additional coordinates needed that have no counter part in conventional Teichmüller theory. For instance, the three holed sphere is not uniquely fixed by its three boundary lenghts anymore. Instead, it has $(N^2 - 1) - 3(N-1)  = (N-2)(N-1)$ internal parameters $s_j$. Thus if we consider a pair of pants decomposition of the Riemann surface $\Sigma_{g,n}$ there are $(N-2)(N-1)$ (which is even for all $N$) parameters $s_j$ associated with each paints. To be more explicit, take $N=3$, then there are three holonomies with $8$ independent components each, this gives 24 parameters. However, we can do an overall conjugation of the three holonomies to eliminate 8 of those and then another 8 of them are fixed by the requirement that the product of the three holonomies must be the identity. This leaves $8$ parameters, $6$ of which can be associated to the three boundaries and so there are two internal parameters. 

Besides the length-coordinates, there are also its conjugate variables, the (generalized) Fenchel-Nielsen twist variables $\tau^{(i)}_j$, which are relevant when gluing three-holed spheres together to form a Riemann surface with higher genus. Together $(\ell_i, \tau_i)$ these are the higher Fenchel-Nielsen coordinates \cite{Goldman1990}. The $\ell_i$ take values in generalisations of the space $\mathcal{R}$ mentioned in \eqref{spaceR}, whereas the twist variables can take any real value. Together with the internal parameters $s_k$ they completely parametrize the Hitchin component. 

To summarize, consider a genus $g$ Riemann surface $\Sigma$ with $n$ geodesic boundaries. We can decompose $\Sigma$ in $-\chi(\Sigma)$ three-holed spheres by cutting along $3g-3+n$ circles. To each three-holed sphere we associate $(N-2)(N-1)$ internal coordinates, to each boundary circle there are $N-1$ length coordinates and finally to each of the $3g-3+n$ circles we cut alongs we have $N-1$ length and twist coordinates. The number of these coordinates adds up to $(2g-2+n)(N^2-1)$, the dimension of the Hitchin component. We can also capture this in the diffeomorphism,
\begin{align} 
\Hit_{g,n}^N \mapsto &\{(\ell_1^{(i)},\dots,\ell_{N-1}^{(i)})\}_{i=1}^n \times \{(\ell_1^{(k)},\dots,\ell_{N-1}^{(k)}),(\tau_1^{(k)},\dots,\tau_{N-1}^{(k)})\}_{k=1}^{3g-3+n}\times \{(s_1^{(j)},\dots,s_{(N-1)(N-2)}^{(j)})\}_{j=1}^{-\chi(\Sigma)}\nonumber\\
&\in \mathcal{R}^n \times (\mathcal{R} \times \mathbb{R}^{N-1})^{3g-3+n} \times \mathbb{R}_+^{-\chi(\Sigma)}\quad, 
\end{align}
with $\mathcal{R}$ the generalisation of \eqref{spaceR} to the $\PSL(N,\mathbb{R})$ case. 

Before moving on to a volume form on this moduli space, let us first study the twist variables in a bit more detail. 

\subsection{Twist coordinates, twist flows and a symplectic measure}

Twist variables in the $\PSL(2,\mathbb{R})$ case have a very geometric interpretation as being the relative angle between two boundaries that are being glued. When quotienting by the mapping class group the twist variables then have a finite range and make the integrals over moduli space finite. For the higher spin theory to make sense\footnote{In the sense that higher genus corrections are finite.}, we would want similar as well. The problem is with this is that one would need to enlarge the mapping class group so that after the quotient the volume is finite. It is fair to say that no concrete proposal for such an extended mapping class group exists. To make some progress, we propose below a natural extension that makes the twist variables $\tau_j^{(i)}$ compact. We will do so in the next subsection, but let us first get some more feeling for what these twist variables mean. 

There are different ways of defining a twist coordinate $\tau_j^{(i)}$. One of them is to consider the length function $\ell_j^{(i)}$ (which is gauge invariant) as a Hamiltonian that generates a Hamiltonian vector field $X_{\ell_j^{(i)}}$. As shown by Goldman \cite{Goldman1986}, this defines a Hamiltonian vector field on $\Hit^N_{g,n}$ and its integral curves are parametrized by the twist coordinate. Thus by flowing along the twist flow one obtains a one-parameter family of flat connections. Geometrically these 'twist-flowed' flat connections are obtained as follows. Consider a surface $\Sigma$ and cut it along a simple closed curve $\g$. This creates two boundary curves $\g_{\pm}$. Do a relative gauge transformation between the two sides by an element of $\PSL(N,\mathbb{R})$ that commutes with the holonomy and then gluing $\g_{\pm}$ back together. By diagonalizing the holonomy around $\g$, the elements that commute with it are generated by the Cartan subalgbra in $\mathfrak{sl}(N,\mathbb{R})$. This algebra has $N-1$ basis elements and so there are $N-1$ different twist flows. 

For instance, consider the cylinder with a marked point on each boundary. The length functions $\ell_j^{(i)}$ are related to the eigenvalues of the holonomy of the gauge field $A$ around a simple closed curve $\g$ of the cylinder (the $A$ cycle). The precise relation depends on a choice of basis for the Cartan subalgebra of $\PSL(N,\mathbb{R})$ and our choice \eqref{defTrumpet} is one such choice for $N=3$. The end result of course (after integration of the length variables) independent of this basis choice. The twist variables are then obtained from the eigenvalues of the holonomy (using the same basis as for the length variables.) around the $B$ cycle, which goes between the marked points on the two boundaries. The $A$ and $B$ cycles are canonical, i.e. have intersection one. Since the length functions $\ell_i^{(j)}$ are Hamiltonians for the twist flow, we can also infer a symplectic form. We have 
\be 
\iota_{V_j}\omega_{\rm Cylinder} = \d \ell_j,\quad V_j = \frac{\partial}{\partial \tau_j}\,.
\ee
Hence, we obtain 
\be \label{omegacyl}
\omega_{\rm Cylinder} = \sum_{j=1}^{N-1} \d \ell_j \wedge \d \tau_j
\ee
In other words, we have the Poisson brackets,
\be 
\{ \ell_i ,\ell_j\} = 0 = \{\tau_i,\tau_j\} = 0,\quad \{\ell_i, \tau_j\} = \delta_{ij}.
\ee
The symplectic measure \eqref{omegacyl} is the straightforward generalisation of Weil-Petersson symplectic form for the $\PSL(2,\mathbb{R})$ case, at least for the cylinder. This argument is heuristic and we have not dealt with the internal coordinates. Luckily, and going back to our choice \eqref{defTrumpet}, including them does not cause a lot more issues and for instance in \cite{WPN3} the full symplectic two-form (on a genus $g$ surface with $n$ boundaries) was worked out in detail for the $N=3$ case,
\be\label{genWP}
\omega_{N=3}(\Sigma_{g,n}) = \sum_{i=1}^{3g-g+n}\sum_{j=1}^{2} \d \ell_j^{(i)} \wedge \d \tau_j^{(i)} + \sum_{j=1}^{2g-2+n} \d s_1^{(j)} \wedge \d s_2^{(j)}\,.
\ee
It is important to note here that \cite{WPN3} also uses the Atiyah-Bott symplectic form \eqref{canonicalSymplecticMeasure} as its starting point to derive this measure.  

\subsection{A proposal for generalized Dehn twists}

Having defined the measure and the twist variables, we can formulate our proposal for the quotient by large (higher) diffeomorphisms. In the $\PSL(2,\mathbb{R})$ case this meant quotienting by the mapping class group of the Riemann surface and for $\PSL(N,\mathbb{R})$ we would want to quotient by an extended version of the mapping class group or rather the group of large higher spin diffeomorphisms. We do not know what that is and instead we will make a proposal for what some of the additional elements of that group need to look like, i.e. what their action is. From the point of view of the BF theory, we will show how we propose they act on the holonomies. This in particular involves how the twist variables change under this action and will be sufficient for our purposes. For the internal coordinates we will have nothing to say, but see \cite{Bekaert:2021sfc, 2018arXiv181211199F, thomas2021higher}.

\begin{figure}
    \centering
    \includegraphics[width=\textwidth]{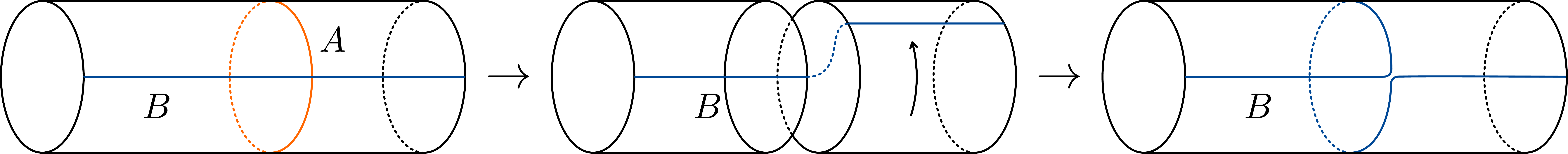}
    \caption{Twist flow and the action of a Dehn twist around the $A$ cycle of the cylinder. The twist flow with parameter $\tau$ is defined by cutting the cylinder at the $A$ cycle, rotating two sides with a relative angle $2\pi \tau/\ell$ and then gluing them back together. Making the angle $2\pi$ or $\tau = \ell$, we make a full rotation and in the process of doing so, the holonomy of along the $B$ cycle (solid blue line) picks up the holonomy around the $A$ cycle. This is why $g_B \to g_A g_B$ (the order does not matter since $g_A$ and $g_B$ commute).}
    \label{fig:DehnTwist}
\end{figure}

Let us briefly recall the the $N=2$ case. The twist flow with twist parameter $\tau = \ell$ is equivalent to a Dehn twist, i.e. a large diffeomorphims. See Fig. \ref{fig:DehnTwist} for a graphical representation of this fact. To quotient by the Dehn twists we therefore  identify a twist flow with twist $\tau = \ell + \tau_0$ with a twist flow with $\tau = \tau_0$. In other words $\tau \sim \tau + \ell$ and hence $\tau$ has a compact range. In terms of holonomies $g_A = \diag(e^{\ell/2},e^{-\ell/2})$ and $g_B = \diag(e^{\tau/2}, e^{-\tau/2})$ around the $A$ and $B$ cycle, the Dehn twist around $A$ acts as $g_A \mapsto g_A$ and $g_B \mapsto g_A g_B$ as can be seen in Fig. \ref{fig:DehnTwist}. Quotienting by this action again tells us that $\tau \sim \tau + \ell$.

In the case $N>2$ there is no immediate geometric picture available for all twist variables, but the aforementioned action in terms of holonomies is a fruitful way to proceed. First, let us consider the usual Dehn twist action. The holonomies are mapped as $g_A \mapsto g_A$ and $g_B \mapsto g_A g_B$, which gives rise to a \emph{single} identification $(\tau_1, \dots , \tau_{N-1}) \sim (\tau_1 + \ell_1, \dots , \tau_{N-1}+\ell_{N-1})$\footnote{This is just the usual (geometric) Dehn twist.}, see also \cite{sun2020volume}. We now propose to generalize this action as follows. Since $g_A$ and $g_B$ are both elements of the Cartan subgroup of $\PSL(N,\mathbb{R})$, we can write them as $g_A = g^{(1)}_A\cdots g^{(N-1)}_A$ and $g_B = g^{(1)}_B\cdots g^{(N-1)}_B$ with $g_{A/B}^{(i)}$ the group element corresponding to the $i$-th generator of the Cartan subalgebra. The action of the generalized Dehn twist we propose is then 
\be\label{genDehntwist}
g_A^{(i)} \mapsto g_A^{(i)},\quad g_B^{(i)} \mapsto g_A^{(i)}g_B^{(i)}, 
\ee
giving rise to the $N-1$ identifications $\tau_i \sim \tau_i + \ell_i$. Notice that this identification is independent of the basis of the Cartan subalgebra. 

Furthermore, since the partition functions we are going to glue only depend on the holonomy around the boundary (so it is independent of the twist variables) and gluing means integrating over the twist and length variables, the basis transformation will cause a change in the partition function because $\mathcal{L}$ changes and the volume form because $\w$ changes. In the end, these are just redifinitions of the $\ell_i$s and since we integrate over them, the basis tranfsormation will not affect the final answer.

This way of presenting the identification smells a lot like we want to geometrize the Cartan directions in $\PSL(N,\mathbb{R})$. One way this can be done is to use the technology of spectral networks \cite{Gaiotto:2012db,Gaiotto:2012rg, Hollands:2013qza, Hollands:2017ahy} and construct the $N$-fold cover of the cylinder (or any Riemann surface) and consider an Abelian connection on it. The holonomy of this Abelian connection around the $A$ and $B$ cycles on the covering manifold will then give rise to a definition of the length and twist variables and the generalized Dehn twists can be viewed as certain Dehn twist of the covering manifold. We will come back to this in section \ref{sec:PSL3Schw}.

\subsection{Wormholes}

With all the technicalities about the gluing measure of two trumpet partition functions out of the way, we can put our proposal to work and consider the generalization of the double trumpet (DT) and its interpretation. Let us consider the $N=3$ case for simplicity. We have, recalling the discussion below \eqref{defTrumpet} and using \eqref{genWP},
\be 
Z_{\rm DT}(\b_1,\b_2) = \int_0^{\infty} \d \ell_1 \int_{-\ell_1/2}^{\ell_1/2} \d \ell_2 \int_{0}^{\ell_1} \d\tau_1 \int_{0}^{|\ell_2|} \d\tau_2 \frac{\g^2}{\pi^2 \b_1 \b_2} \exp\left( - \frac{\g (\b_1 + \b_2)}{2 \b_1 \b_2} \left(\frac{\ell_1^2}{4} + \frac{\ell_2^2}{3}\right) \right),
\ee
where we have put $|\ell_2|$ for the range of $\tau_2$ as it can be negative, but the integral over the compact $\tau_2$ direction is positive. The integrals over the twist variables give a measure factor $\ell_1 |\ell_2|$ and performing the remaining integrals over $\ell_1$ and $\ell_2$ gives the following expression for the $N=3$ double trumpet,
\be 
Z_{\rm DT}(\b_1,\b_2) = \frac{6}{\pi^2} \frac{\b_1 \b_2}{(\b_1 + \b_2)^2}.
\ee
If we now perform the analytic continuation to Lorentzian signature (i.e. to calculate the spectral form factor) $\b_1 \to \b + i T$ and $\b_2 \to \b - i T$ and take the large $T$ limit, we obtain
\be \label{cylinderResultlargeT}
Z_{\rm DT}(\b + i T,\b - i T) = \frac{3}{2\pi^2} \frac{T^2}{\b^2} + \dots,
\ee
with the dots representing lower order in $T$ contributions. We see that instead of linear in $T$ we get quadratic in $T$ behaviour at late times. This behaviour we will modify the dual matrix model in an essential way.

It is not too complicated to see that in the case of $\PSL(N,\mathbb{R})$ with the definition for the trumpet as mentioned and the gluing measure from the above discussion, the general $N$ answer has the form
\be \label{SFFHS}
Z_{\rm DT}(\b_1,\b_2) = c_N \left(\frac{\sqrt{\b_1\b_2}}{\b_1 + \b_2}\right)^{N-1}\quad\Rightarrow\quad Z_{\rm DT}(\b + i T, \b - i T) = c_N \left(1 + \frac{T^2}{\b^2}\right)^{\frac{N-1}{2}}
\ee
with $c_N$ a $N$ dependent coefficient. We thus see that at late times the double trumpet contributes goes like $T^{N-1}$. This was also seen in \cite{Das:2021ipx} in $3d$, but relied on a particular assumption about the gluing measure. Here we tried to argued for the gluing measure from first principles.

The large $T$ limit also differs slightly from the usual case in another way. The dip time $T_{\rm dip}$ (the time at which the cylinder and two disks become comparable) depends on $N$. To see this notice that the disconnected contribution goes like $e^{2\S} T^{1-N^2}$ and so by comparing to \eqref{SFFHS}, $T_{\rm dip}$ scales like
\be 
T_{\rm dip} \sim e^{\frac{2\S}{(N+2)(N-1)}}.
\ee
Hence for $N=3$, we want $T \gg e^{\S/5}$.\footnote{Notice that the ramp starts earlier when $N$ increases. This will be interesting when we study the large $N$ limit in section \ref{sec:discussion}.} Just as in the JT case, $T$ cannot be too big either and must be smaller then the Heisenberg time $T_{\rm Heis}$. This we can only discuss when we have a proposal for the boundary theory, which will discuss now. 

\section{A matrix model dual}\label{sec:matrixmodel}

The gravitational calculation of the wormhole contribution to the spectral form factor indicated a serious deviation from conventional hermitian one-matrix models. Here we propose a simple multi-matrix model that reproduces the enhanced ramp behaviour. 

\subsection{Commuting multimatrix models}

The higher spin theories that we have considered not only have the Hamiltonian $H$ as a conserved boundary charge, but for each higher spin field there is conserved charge as well. Including $H$ there are thus $N-1$ converved charges $\mathcal{Q}_i$ on the boundary with $i = 2,\dots,N$ and $\mathcal{Q}_2 = H$. These conserved charges by definition commute with $H$ but also with each other.\footnote{Mutual commutation of the charges is a bit subtle and might require a simple (local) redefinition of the spin$-N$ charges one gets on the boundary through the procedure outlined in section \ref{sec:setup}. This subtlety arises for $N\geq 4$. For instance in the $N=4$ case, the usual $\Wc_4$-algebra has non-commuting spin$-3$ and spin$-4$ zero modes \cite{Blumenhagen:1990jv}, but a simple redefinition of the spin$-4$ field ensures that they commute \cite{Niedermaier:1991sq, Niedermaier1992}. } This allows us to define a generalized Gibbs ensemble of the form 
\be \label{generalizedGibbs}
Z(\mu_2=\b,\m_3,\dots, \mu_N) = \Tr e^{- \sum_{k=2}^{N-1} \mu_k \mathcal{Q}_k}\,.
\ee
These particular ensembles have been discussed in the past in the integrability literature \cite{GGE1, PhysRevLett.98.050405} and for us it suggests that the matrix model dual is an ensemble of not just $H$ but of $N-1$ $L$ by $L$ commuting hermitian matrices. The matrix model we want to consider is thus,
\be \label{matrixmodelHS}
\mathcal{Z} = \int \d \mathcal{Q}_2 \dots \d \mathcal{Q}_N e^{-L \Tr V(Q_i)}
\ee
where we integrate over the space of $N-1$ commuting hermitian matrices. This means we can simultaneously diagonalize all the matrices and writing this matrix model purely in terms of the eigenvalues of the matrices involved. \footnote{This matrix model should be constrasted with the one discussed in \cite{Kapec:2019ecr, Iliesiu:2019lfc}, where one has a global symmetry $G$ on the boundary, that is realized in the bulk as a $2d$ BF theory. There, one can assume that within each representation sector labelled by $R$ the Hamiltonians $H^{(R)}$ are independent random matrices. For us this is not the case. First, we do not have a global symmetry on the boundary, i.e. our boundary conditions are not that of a particle on the $PSL(3,\mathbb{R})$ group manifold. Second, the spin two and higher spin fields interact in the bulk, unlike when you have $2d$ BF in the bulk. The partition function in a fixed representation is just a product of the gravity and gauge theory partition functions. Furthermore, since the gauge theory partition function just depends on the sum of the boundary lengths, you cannot alter the RMT behaviour, which we did see happening when higher spin fields are included. It could be that when one has a 2d gauge theory in the bulk, but with a non-trivial coupling to the dilaton, which makes the gauge variables interact with gravity, this could change. We thank Joaquin Turiaci for discussions on this. } To accomplish this more rigorously one can consider starting from an integral over \emph{all} hermitian matrices and insert a product of delta functions $\prod_{i<j}\delta(U_i U_j^{-1})$ to force all the unitaries to be equal so that the matrices are simultaneously diagonalized. Note that we assume here that the potential is single trace and unitary invariant. Denoting the matrices collectively by $\mathcal{Q}^a$, we have $\mathcal{Q}_a = U \L_a U^{\dagger}$. In a ordinary one-matrix model the measure coming from diagonalization is a Vandermonde determinant. In our case it will be slightly different and supply the crucial ingredient in explaining a modified ramp behaviour. To compute the measure we consider
\be 
\d s^2 = \Tr \left(\d \mathcal{Q}^a \d \mathcal{Q}_a \right)
\ee
We have $\d \mathcal{Q}^a = U (\d \L^a + [U^{\dagger}\d U, \L^a])U^{\dagger}$ and after a bit of algebra (see also \cite{Berenstein:2005jq, Filev:2014jxa}) we find
\be 
\d s^2 = \sum_{i=1}^L \d \l_i^a \d \l_{a,i} + \sum_{i\neq j} \left(\sum_a (\l^a_i - \l^a_i)^2\right) \theta_{ij}\theta_{ij}^{\dagger}
\ee
with $\theta_{ij} = U_{ik}\d U^{\dagger}_{kj}$. After doing the (trivial) unitary integral, the matrix integral becomes
\be 
\mathcal{Z} = \int \d \l_i^1\cdots \d \l_i^{N-1} \prod_{i<j} \left(\sum_a (\l^a_i - \l^a_i)^2\right) e^{-L \Tr V(\L^a)}\, .
\ee
One immediately sees that due to the presence of $N-1$ commuting matrices the object in place of the standard Vandermonde determinant has a similar repulsive behaviour but now depends on the total distance between two different eigenvalues. In fact, we are now not dealing with a Dyson gas of $L$ particles in $d=1$ dimension, but in $d = N-1$ dimensions. We will often denote by $\l$ or $\l_i$ a $N-1$ dimensional vector of eigenvalues.

\begin{figure}
    \centering
    \includegraphics[scale=0.35]{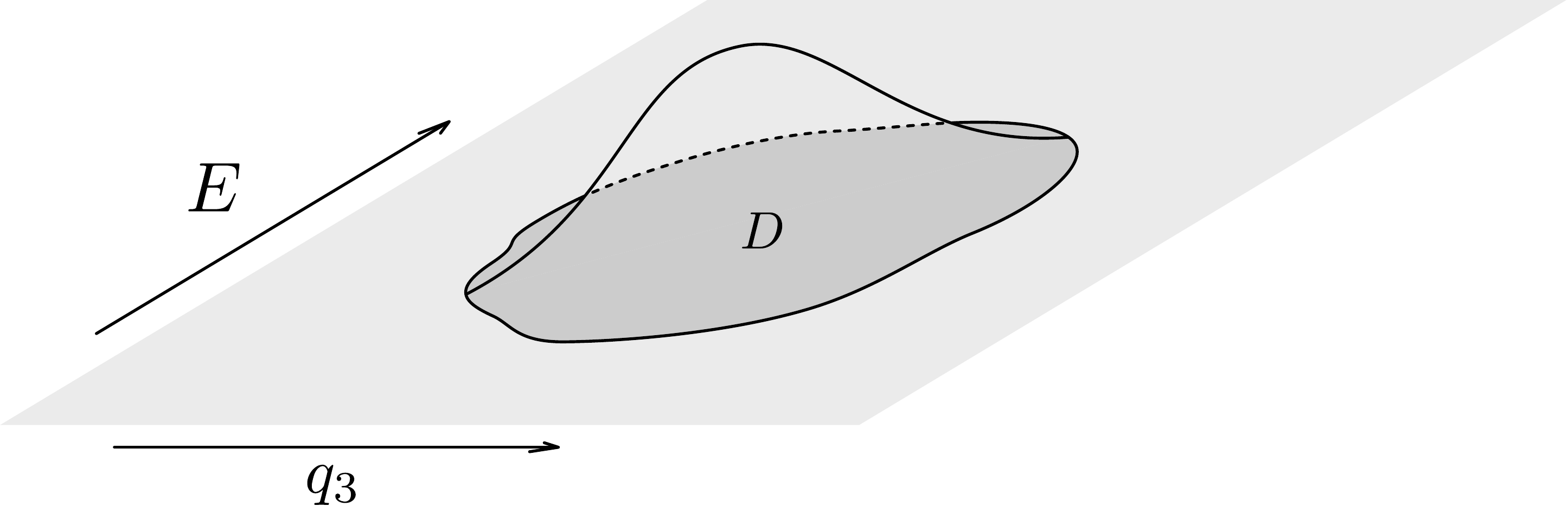}
    \caption{Continuous spectral density $\rho(E,q_3)$ for the $N=3$ theory. The distribution is two dimensional and constraint to some region $D$ in the $(E,q_3)$ plane. For $N$ is odd, the spectral density can go smoothly to zero at the boundary of $D$ if the potential is of high enough degree.}
    \label{fig:2dSpectraldensity}
\end{figure}

Let us analyze this model in a bit more detail. First, it is instructive to study the saddlepoints of the matrix integral in the large $L$ limit. This will allow us to understand what types of potentials we should be dealing with in order to get the density of states we found in the gravitational calculation. At finite $L$ we have a discrete distrubtion of eigenvalues,
\be 
\rho(\l) = \sum_{i=1}^L \delta(\l_1-\l_{1,i})\cdots \delta(\l_{N-1}-\l_{N-1,i}) = \sum_{i=1}^L \delta^{(N-1)}(\l - \l_i)
\ee
In the large $L$ limit it is convenient to treat the distribution of the $L$ particles in $d=N-1$ dimensions as continuous and introduce a density $\rho(\l)$, see Fig. \ref{fig:2dSpectraldensity} for an example for $N=3$, that is constraint to integrate to $L$ over some domain $D \in \mathbb{R}^{N-1}$,
\be \label{constraint}
\int_D \d^{N-1}\l \, \rho(\l) = L\, .
\ee
In the matrix integral this can be accomplished by using a Lagrange multiplier. We also consider only non-negative $\rho$. The on-shell equations then take the form,
\be \label{MatrixModelSaddles}
\frac{L}{2} \partial_{\l^a} V(\l) = \int \d^{N-1}\l' \, \rho(\l') \frac{\l^a - \l'^a}{|\l - \l'|^2}
\ee
As noted in \cite{Berenstein:2005aa} when $N$ is odd, we have $(\partial_{\l^a}\partial_{\l_a})^{(N-1)/2} \log |\l - \l'|^2 \sim \delta^{(N-1)}(\l - \l')$ and so this equation can have solutions for $\rho$ that vanish smoothly near the edge of $D$, whenever $V$ is of sufficiently high degree, i.e. $\rho$ has no root singularity close to the edge of $D$ such as we would have for instance in a one-matrix model. Concretely, for $N=3$, by taking another derivative of \eqref{MatrixModelSaddles} with respect to $\l^a$ and summing over $a$ we get 
\be 
\frac{L}{2} \nabla^2 V(\l) = 2\pi \rho(\l).
\ee
This equation (together with \eqref{constraint}) are trivial to solve and for instance in the case of a gaussian potential, $\rho$ is a constant in some domain $D$ in $\mathbb{R}^2$, but for higher degree $V$ one can engineer $\rho$ to vanish as a power law at the edge of $D$. 

The difference between even and odd $N$ that we see is also reflected by the gravitational calculation. For $N$ odd we found that the spectral density goes smoothly to zero at zero energy instead of like a root, whereas for $N$ even, the distribution $\rho$ is not smooth function and one needs to solve \eqref{MatrixModelSaddles} through other methods. Again, for the $N=3$, the gravitational calculation suggests that near the edge of the spectral density ($E=0$) goes to zero as $E^3$. By taking the potential $V$ to be at least fifth order (in the eigenvalue corresponding to the energy), we can accomplish such behaviour. However, since we have not done the full gravitational calculation with non-zero chemical potential for the higher spin charge, we can only fix the potential partially. Furthermore, we have assumed here a particular double scaling procedure in which we take $L$ to infinity and scale towards the edge such that we keep a finite density. Nevertheless, the situation is a bit similar to ordinary JT gravity. There the potential is also not known, but defined in a limiting procedure using the minimal string theories. Here we do not have this alternative perspective, but it is clear that from the equations above, we could in principle reverse engineer the potential by matching with the density of states of the gravitational calculation. We also note here that we do not expect the potential $V$ to have any symmetries, such as rotational invariance. A difference with JT is that since we have multiple matrices on the boundary, how do we know what matrix is the boundary Hamiltonian? This should be determined by the specific double scaling limit of the multi-matrix model, since we want the energy variable to take all positive values. It would be valuable to understand this in more detail. 

\subsection{Fluctuations}

The saddle point solutions are interesting, but in some sense we would fix them to get the correct density of states obtained from the gravitational calculation. It would be considered an input to the matrix model definition, just like in JT gravity \cite{Saad:2019lba}. The real test of our proposal is a more intrinsic property, namely its eigenvalue repulsion. In JT gravity case this leads to a linear-in-time behaviour of the spectral form factor $\braket{|Z(\b+\i T)|^2}$. In our case we are studying a generalization of the spectral form factor,
\be \label{SFFgen}
g(\b,T,\{\mu_i,\xi_i\}) = \braket{|Z(\b + \i T,\{\mu_i + \i \xi_i\})|^2},
\ee
where the average is taking in the matrix model \eqref{matrixmodelHS}. On the gravity side we found that for $T\gg T_{\rm dip}$ it behaves as $T^{N-1}$ and our matrix model should reproduce that. Let us now show that this is the case by studying the spectral two-point function.

The leading order connected component of the spectral two-point function follows from the quadratic fluctuations around the saddle and solely comes from the generalized Vandermonde, i.e. we can focus on the term in the matrix model action $I$ that is quadratic in $\rho$ (see \cite{Cotler:2016fpe} for the calculation in ordinary RMT),
\be 
I \supset I_{(2)} = -\int \d^{N-1}\l_1 \d^{N-1} \l_2 \delta \rho(\l_1)\delta \rho(\l_2) \log |\l_1 - \l_2|
\ee
This is a non-local action, but can be brought into a local one by doing a Fourier transform. Let us write $\delta \rho(\l)$ as, 
\be 
\delta \rho(\l) = \int \frac{\d^{N-1} s}{(2\pi)^{N-1}}\; e^{\i s^a\l_a} \delta \rho(s),
\ee
and go to coordinates $\l_{+}^a = \l_1^a + \l_2^a$ and $\l^a = \l_1^a - \l_2^a$. We then have 
\be 
I_{(2)} = -\int \frac{\d^{N-1}s}{(2\pi)^{N-1}} \delta\rho(s) G(s) \delta\rho(-s)
\ee
with $G(s^a)$ given by
\be 
G(s) = \int \d^{N-1} \l \; e^{\i s^a \l_a} \log |\l|
\ee
The two-point function of the fluctuations, which gives the generalized ramp behaviour, is 
\be 
\braket{\delta\rho(s^a)\delta\rho(-s^a)} = -\frac{1}{2}(2\pi)^{N-1}G(s^a)^{-1}.
\ee
To evaluate $G(s)$ we can go to spherical coordinates such that $s^a\l_a = |s||\l| \cos \phi = s \l \cos\phi$ and we obtain
\be 
G(s) = {\rm Vol}(S^{N-3}) \int_0^{\infty} \d \l \int_0^\pi \d \phi \sin^{N-3}\phi\; e^{\i s \l \cos \phi} \l^{N-2} \log \l
\ee
Using 
\be 
\int_0^\pi \d \phi \, e^{\i s \l \cos \phi} \sin^{N-3}\phi = \pi^{1/2} \G(N/2 - 1) \frac{J_{\frac{N-3}{2}}(s\l)}{(s\l/2)^{\frac{N-3}{2}}}
\ee
with $J_\nu$ the Bessel function of the first kind, the radial $\l$ integral is a bit tricky, but can then be done. We find\footnote{One can for instance calculate the integral for arbitrary $N$ or one can also multiply $G(s)$ with suitable powers of $|s|^2$ and converting those to Laplacians (in $\lambda$) that act on the logarithm. For instance for $N=3$, since $\log|\l|$ is the Green function on the plane, we immediately get the equation $-|s|^2 G(s) = 2\pi$. This also has a delta function $\delta(|s|^2)$ as a solution but that distribution is identically zero because when integrating against a smooth test function the measure gives a factor $|s|^2$.}
\be 
G(s) = -2^{N-2}\pi^{\frac{N-1}{2}} \G\left(\frac{N-1}{2}\right) \frac{1}{|s|^{N-1}} 
\ee
And hence, 
\be \label{fluctTwo}
\braket{\delta\rho(s)\delta\rho(-s)} = \frac{\pi^{\frac{N-1}{2}}}{\G\left(\frac{N-1}{2}\right)}|s|^{N-1}.
\ee
To see this gives the generalized ramp, notice that $s$ is a vector of times, including `times' $\xi_i$ that come from analytic continuation of the higher spin chemcial potentials $\mu_i \to \mu_i \pm \i \xi_i$ (in analogy with $\beta \to \beta \pm \i T$). Taking $T$ much larger than the other 'times', we get 
\be \label{fluctTwoT}
\braket{\delta\rho(s^a)\delta\rho(-s^a)} \sim T^{N-1},
\ee
or when translating to the two-point function of the generalized spectral form factor,
\be 
\braket{|Z(\i T, \{\i \xi_i\}|^2} \sim T^{N-1},\quad T \gg e^{\frac{2\S}{(N+2)(N-1)}},
\ee
as desired.\footnote{For the $N=3$ case, where we have two commuting matrices, we can put them in one complex matrix $M$, which should commute with its hermitian conjugate. Such matrix models are known as normal matrix models. In fact, the Vandermonde there is equal to the one of just complex matrices and so will give rise to a $T^2$ behaviour as well. This is the Ginibre ensemble if the potential is Gaussian, see \cite{ProsenGinibre} for some exact formulas. In that respect the $N=3$ case is in fact special and to understand that we need commuting matrices we need to study the general $N$ case. From the way we motivated the matrix model, the $N=3$ is also special in the sense that the spin$-3$ charge is conserved, so commutes with the other matrix, the Hamiltonian, but for $N>3$ we also get the fact that the higher spin charges commute.} 
It is worthwhile to calculate the density correlator in the eigenvalue space. Just as for the leading calculation of the density, there is again a difference between even and odd $N$. For $N$ odd, so $N=2m+1$, we have 
\be 
\braket{\delta \rho(\l_1)\delta \rho(\l_2)} = \frac{\pi^m}{\G(m)} \int \frac{d^{2m}s}{(2\pi)^{2m}} |s|^{2m} e^{i s\cdot (\l_1 - \l_2)},
\ee
but this can be written as a certain power of the Laplacian acting on the $2m$-dimensional delta function and so we arrive at
\be 
\braket{\delta \rho(\l_1)\delta \rho(\l_2)} = \frac{\pi^m}{\G(m)}(-\nabla_{\l_1}^2)^m \delta^{(2m)}(\l_1 - \l_2)
\ee
This means that even though the spectral correlation in eigenvalue space is ultra local, it can still give rise to a power law behaviour at late times $T$. For even $N$ we cannot use this trick and we get a power law behaviour in eigenvalue space as well. For instance, for $N=4$ we obtain
\be 
\braket{\delta \rho(\l_1)\delta \rho(\l_2)} = \frac{24}{\pi} \frac{1}{|\l_1 - \l_2|^6},
\ee
which can be checked gives $2\pi |s|^3$ as we found in \eqref{fluctTwo}.

\begin{figure}
    \centering
    \includegraphics[scale=0.4]{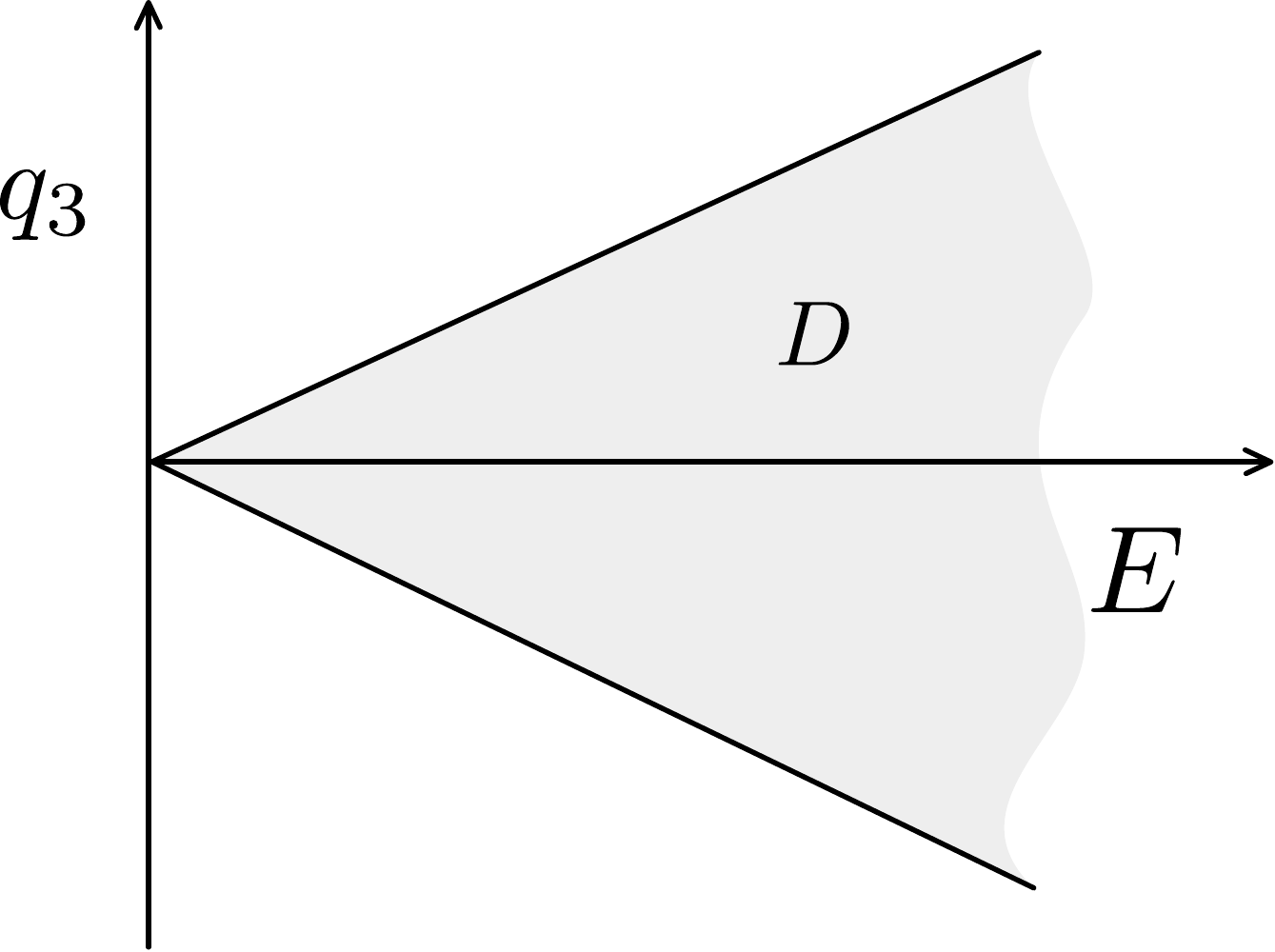}
    \caption{Wedge shaped region for the $N=3$ theory after double scaling in order to match with the gravitational result.}
    \label{fig:DoubleScaling}
\end{figure}

Furthermore, the correlator of fluctuations is universal and does not depend on the details of the matrix model potential $V$. Consquently the $T^{N-1}$ behaviour is universal. However, in contrast to the ordinary JT gravity, the coefficient of the $T^{N-1}$ in the generalized spectral form factor in the $N>2$ theory is not quite universal. Recall that for the $N=2$ theory the coefficient was $1/(2\pi \b)$ and its universality follows from the Schwarzian having positive energy. In the $N>2$ theory we do not only integrate over energy but also over the value of the additional higher spin charges. In energy and higher spin charge basis, the generalized spectral form factor at $\mu_i = 0$ can be written as (at large $T$)
\be 
\braket{|Z(\b + \i T,\{\i \xi\})|^2} = T^{N-1}\left(\frac{\pi^{\frac{N-1}{2}}}{\G\left(\frac{N-1}{2}\right)}\int_D \frac{\d E \d^{N-2} q_i}{(2\pi)^{N-1}} e^{-\b E}\right),
\ee
where we used \eqref{fluctTwoT}. The region $D$ is the region where the spectral density of the matrix model saddlepoint solution has non-zero support. We want the term in round brackets to match the gravitational expression \eqref{cylinderResultlargeT}. This puts some mild constraints on $D$. For instance in the $N=3$ theory we want 
\be 
\frac{3}{2\pi^2 \b^2} = \frac{1}{4\pi} \int_0^{\infty} \d E \int_{\tilde{D}} \d q \, e^{-\b E}
\ee
where we assumed the integral over $E$ covers the positive real line and the spin-3 charge takes values in $\tilde{D}$, which could depend on $E$. To the level of approximations we are working here this means that we should have 
\be \label{constraintSpin3}
\int_{\tilde{D}} \d q = \frac{6 E}{\pi}.
\ee
This means that the region $\tilde{D}$ should depend linearly on $E$ and so this breaks the universality that we had for the $N=2$ theory, but only slightly so. This also depends on how one performs the double scaling in the $N>2$ theories. Here it seems we want to scale towards the edge of the domain $D$ in the $E$ direction by simultaneously squeezing the $q_3$ direction such that \eqref{constraintSpin3} holds, see Fig. \ref{fig:DoubleScaling}. It would be interesting to make this double scaling more precise.
Perhaps this bound on the range of $\tilde{D}$ is related to a unitarity bound. For instance a similar bound on the spin$-3$ charge, but in the context of 2d CFT was found in \cite{Afkhami-Jeddi:2017idc}. 

To close off this section, we would like to make a few comments about the time at which non-perturbative effects in the matrix model should kick in.\footnote{We thank Steve Shenker for discussions on this.} This is usually related to the time at which we probe energies of the size of the average level spacing of the system. This time is known as the Heisenberg time and let us try to estimate it in our model. Let us consider our large $L$ matrix model with some potential $V$. The density of eigenvalues is confined in some region $D$ and its integral is $L$. The domain $D$ depends on some parameters of the model, but not on $L$. Thus the average distance between two eigenvalues is determined by filling up $D$ with small balls with radius $\e$ such that
\be 
L = \frac{\#}{\e^{N-1}}
\ee
with $\#$ some numerical constant independent of $L$. Thus the average radius of the ball each eigenvalue occupies is $\e \sim L^{1/(1-N)}$. The Heisenberg time is the inverse of this, so 
\be 
T_{\rm Heis} \sim L^{\frac{1}{N-1}}
\ee
For $N=3$ this was also found in \cite{ProsenGinibre} for the Ginibre ensemble. In the double scaling limit we then imagine $L \sim e^{\S}$ and so the regime of validity of the cylinder contribution is $e^{2\S/(N-1)(N+2)}\ll T \ll e^{\S/(N-1)}$. 

\section{Towards a geometric description}\label{sec:PSL3Schw}

In this section we change gears a little bit and will not further discuss the matrix model, but instead try to find a more appealing geometric interpretation of the higher spin theories and of the generalizations of the Dehn twists proposed in section \ref{sec:gluing}. Usually this is difficult due to the non-linear nature of the $\Wc_N$ algebra\footnote{Here with non-linear we mean that the algebra does not just generate transformations generated by vector fields, but also by higher-order tensor fields containing more than just one derivative.}, but luckily there is a neat way of linearizing the algebra, as we will show below. This amounts to introducing $N-2$ additional (auxiliary) coordinates on the boundary such that the $\Wc_N$ transformations lift to a set of ordinary diffeomorphisms in the enlarged space and the $N-2$ higher spin conserved charges generate translations in these new directions. 

This geometrification of the $\Wc_N$ symmetries also allows us to construct a generalization of the Schwarzian for $\PSL(3,\mathbb{R})$ in case of the disk and trumpet. When expanding around the saddle points we find exactly the same quadratic action as the one discussed in \ref{sec:disktrumpet}. We will further comment on what the fields $F_i$ and their integration space mentioned in section \ref{sec:setup} is. 

Since this section involves some formalism, let us highlight the main steps,
\begin{enumerate}
    \item We first introduce additional variables that can be thought of as 'additional' time coordinates through deformations that leave the holonomy of the gauge field invariant.
    \item We use these coordinates to construct a locally flat manifold from which we can extract the equations determining the flat connection $A_u$ by picking a certain gauge for the Christoffel symbols. 
    \item Using diffeomorphisms that leave this gauge invariant, we construct an explicit form of the Schwarzian derivative relevant for the $\PSL(3,\mathbb{R})$ disk and trumpet and their integration spaces.
\end{enumerate}

At the end of this section we use the additional boundary times to construct a bulk that geometrizes some of the higher spin symmetries that are otherwise difficult to understand. For trumpet in the $N=3$ theory we put forth a geometric description in terms of a three dimensional in which the two symmetries of the trumpet are ordinary isometries. We use this description also to find concrete evidence for the existence of the generalized mapping class group elements proposed in section \ref{sec:gluing}. 

\subsection{We need more time}

In the BF gauge theory, the main objective is finding all flat connections on a given manifold with particular boundary conditions. For instance, for the $N=3$ theory on the disk, we need to solve the equation $A_u = g^{-1}\partial_u g$ with $g$ having trivial monodromy as discussed in section \ref{sec:disktrumpet}. To solve this equation it is convenient to write $g$ as
\be \label{gconn}
g = \begin{pmatrix}
\psi_1 & \frac{\psi_1'}{\sqrt{2}} & \frac{1}{2}(\psi_1'' + 2\mathcal{L} \psi_1)\\  
\psi_2 & \frac{\psi_2'}{\sqrt{2}} & \frac{1}{2}(\psi_2'' + 2\mathcal{L} \psi_2)\\
\psi_3 & \frac{\psi_3'}{\sqrt{2}} & \frac{1}{2}(\psi_3'' + 2\mathcal{L} \psi_3)\\
\end{pmatrix}
\ee
with $\psi_i$ linear independent (so the determinant can be normalized to one) solutions of $L \psi_i = 0$ with 
\be \label{L}
L = \partial_u^3 + 4 \mathcal{L} \partial_u + (2\mathcal{L}' - 8 \mathcal{W})\,.
\ee
From this way of writing the solutions it becomes evident that the holonomy of the gauge field around the boundary is equivalent to the monodromy of the solution vector $\psi = (\psi_1,\psi_2,\psi_3)^T$ around the boundary. Each set of $\psi_i$ that has trivial monodromy gives rise to a valid flat connection $A$ on the disk. These functions $\psi_i$ actually give rise a set of coordinates $(\psi_1/\psi_3, \psi_2/\psi_3)$ on $\mathbb{RP}^2$. The way to see this is by noticing that we can multiply $g$ on the left by a constant $h \in \PSL(3,\mathbb{R})$ and obtain the same connection $A_u$. This has the effect of taking linear combinations of $\psi_i$ (which is still a solution due to the linearity of $L$) and so the ratios $s_i = \psi_i/\psi_3$ naturally transform as fractional linear transformations,
\be 
s_i \to \frac{a_1^i s_1 + a_2^i s_2 + a_3^i}{a_1^3 s_1 + a_2^3 s_2 + a_3^3}
\ee
with $a_i^j$ the elements of an $\PSL(3,\mathbb{R})$ matrix. This generalizes the Mobius transformations acting on the real projective line in the $N=2$ case. This projective action will become important later when we analyze the geometry after introducing the aforementioned additional boundary coordinates. 

To introduce such coordinates we want to find deformations that result in different $g$, but keep the holonomy of $A_u$ the same. The holonomy should thus not dependent on such additional coordinates. As explained for instance in \cite{Govindarajan:1994wm}, there are such deformations, in fact well-known deformations, namely the isomonodromic deformations of the generalized KdV hierarchy. These deformations are flow equations for the operator $L$ given by
\be \label{LKdV}
\partial_{t_p} L = [(L^{p/N})_+,L],
\ee
where $(\dots)_+$ means taking the differential operator part and $p = 1,\dots,N-1$. The KdV times $t_p$ are the naturally additional boundary coordinates that lift the non-linear $\Wc_N$ transformations to ordinary diffeomorphisms. We will see how this works momentarily. First, in order to understand why these deformations preserve the monodromy, act with $\partial_{t_p}$ on $L\psi = 0$ to obtain
\be 
L \left(\partial_{t_p}\psi - (L^{p/N})_+\psi \right) = 0
\ee
This means that the term in brackets must also be annihilated by $L$. Thus the term in round brackets is a $u$ independent matrix $\L$ times $\psi$. To bring the monodromy matrix $M$ in the discussion, we transport $\psi$ around the thermal circle. This gives a flow equation for $M$,
\be 
\partial_{t_p} M = [\L,M]\, ,
\ee
which tells us that $M$ does not change conjugacy class under the flow along $t_p$. However, we wanted $\partial_{t_p} M = 0$, i.e. isomonodromic deformations. Thus, $\L$ needs to commute with $M$. For the case of general $M$, this means $\L$ is proportional to the identity and in fact we can pick the proportionality constant to be zero, which amounts to identifying $t_1$ with $u$. \footnote{There are other choices of $\L$ that also gives rise to isomonodromic deformations, but this choice will be convenient for us. If the base space was a complex surface, i.e. instead of just $u$ we would have a complex coordinate $z$, the monodromy matrix takes different forms for different cycles on the surface and so it makes sense to set $\L$ proportional to the identity and then pick the proportionality constant to vanish. Here we have only the boundary circle, so one could allow for other $\L$ as well.} At any rate, when $\L = 0$, the flow of the solutions is given by 
\be 
\partial_{t_p}\Psi_i = (L^{p/N})_+ \Psi_i,\quad \Psi_i(t_1,\dots,t_N)|_{t_i = 0} = \psi_i(u).
\ee
As an example, let us consider $N=3$. We have $(L^{1/3})_+ = \partial_u$, so $\partial_{t_1} = \partial_u$ or $t_1 = u + \text{const.}$, and $(L^{2/3})_+ = \partial_{t_1}^2 + \frac{8}{3}\mathcal{L}$, so the non-trivial flow equation is 
\be
\partial_{t_2}\Psi_i(t_1,t_2) = \partial_{t_1}^2 \Psi_i(t_1,t_2) + \frac{8}{3}\mathcal{L} \Psi_i(t_1,t_2).
\ee
These $\Psi_i$ still solve $L \Psi_i = 0$ and so again define a flat connection on the disk.  This suggests that it is more convenient to enlarge the dimensionality of the boundary by $N-2$ additional coordinates. Let us denote this geometry by $M_{N-1}$. From this higher-dimensional perspective there is actually a nice way of obtain the equations \eqref{L} and \eqref{LKdV} from a flatness conditions of the $N-1$ dimensional space \cite{Gomis:1994rz}. Consider the equations
\be \label{integrabilityEqs}
\nabla_a \nabla_b \Psi_i(t_1,\dots,t_{N-1}) = 0,\quad a,b = 1,\dots, N-1
\ee
with the covariant derivative is defined so that $\Psi_i$ has weight $1/N$ under diffeomorphisms of $M_{N-1}$,\footnote{This should be contrasted with the weight of $\psi_i$ under diffeomorphisms of $u$ only, which is $-(N-1)/2$. The $1/N$ has been chosen so that when we reduce from $M_{N-1}$ back to the boundary circle, we get the correct weight \cite{Gomis:1994rz}. }
\be 
\nabla_a \Psi_i = \partial_a \Psi_i + \frac{1}{N} \Gamma^b_{ba} \Psi_i \,.
\ee
We now want to find conditions on the Christoffel symbols $\G^a_{bc}$ such that there $N$ linearly independent solutions of \eqref{integrabilityEqs}. It is clear that when $M_{N-1}$ was flat space, we would indeed get $N$ independent solutions (the $N$th solution is simply the constant solution). In fact, the solution for \eqref{integrabilityEqs} is just a diffeomorphism of that, i.e. there are only $N$ linear independent solutions if the Riemann tensor vanishes and the Christoffel symbols are symmetric. This follows from considering the integrability conditions $[\nabla_a,\nabla_b]\Psi_i = 0$ and and $[\nabla_a,\nabla_b]\nabla_c\Psi_i=0$ \cite{Gomis:1994rz}. 

Now, the point of \cite{Gomis:1994rz} is to consider a gauge for the Christoffel symbols in which we reproduce \eqref{L} and \eqref{LKdV}. To be explicit consider $N=3$ so that we have two coordinates $(t_1=u,t_2)$. Following \cite{Gomis:1994rz} we impose the gauge condition $\G^2_{11} = 1$ and $\G^1_{11} = 2 \G^2_{12}$. There are then three equations \eqref{integrabilityEqs} to be solved but by using the vanishing of the Riemann tensor only two are non-trivial and read,
\be\label{reduction}
\partial_1^3 \Psi_i + 4 \mathcal{L} \partial_1 \Psi_i + (2 \mathcal{L}' - 8 \mathcal{W})\Psi_i = 0,\quad \partial_2 \Psi_i = \partial_1^2 \Psi_i + \frac{8}{3}\mathcal{L}\Psi_i,
\ee
which are precisely the equations we wanted to reproduce. Here 
\begin{align}
\mathcal{L}(t_1,t_2) = \frac{1}{4}(\G^2_{22}(t_1,t_2) - 2 \G^1_{12}(t_1,t_2))\label{LGamma}\\
\mathcal{W}(t_1,t_2) = \frac{1}{8}(\G^1_{22}(t_1,t_2) + \frac{2}{3} \partial_1 \mathcal{L}(t_1,t_2)) \label{WGamma}
\end{align}
Thus we see that by geometrizing the solutions $\Psi_i$, we also got a nice expression for the stress tensor $\mathcal{L}$ and spin-3 field $\mathcal{W}$.

To summarize, we have introduced additional coordinates $t_p$ through flow equations that leave the holonomy of $A_u$ invariant. We then reinterpreted these flow equations as coming from flatness conditions on some higher dimensional space $M_{N-1}$ and wrote both $\mathcal{L}$ and $\mathcal{W}$ in terms of geometric objects of $M_{N-1}$.

\subsection{$\PSL(3,\mathbb{R})$ Schwarzians}

Using the form \eqref{LGamma} we can now find the correct $\PSL(3,\mathbb{R})$ generalization of the Schwarzian theory. Before proceeding, let us recall some of the intuition for the ordinary Schwarzian theory. The action can be obtained from a geometric construction, i.e. considering the extrinsic curvature of a curve in the hyperbolic disk. There one considers a path integral of the Schwarzian action over all curves that are diffeomorphisms of the `round' circle, i.e. one integrates over all boundary `wiggles'. These are different cutouts of the hyperbolic disk. 

The analogue of these diffeomorphisms in our case are the gauge transformations that leave the asymptotic form of the gauge field \eqref{AAdS2bc} invariant. However, since the $\Wc$-transformations are non-linear in terms of just the $u$ coordinate it is hard to make this analogy precise non-linearly (on the linear level this was done in section \ref{sec:setup}). Fortunately, the additional coordinate $t_2$ simplifies this, because it allows us to talk about the gauge transformations that leave \eqref{AAdS2bc} invariant as ordinary diffeomorphisms in the locally flat manifold $M_2$ parametrized by $(t_1,t_2)$. The catch is that since we already picked a gauge for the Christoffel symbols, these diffeomorphisms need to preserve that condition and cannot be arbitrary. This relates the $t_2$ derivatives of the diffeomorphisms to derivatives wrt to $t_1$. Denoting the diffeomorphisms by $t_i \to F_i(t_1,t_2)$ the constraints can be derived rather easily and take the form \cite{Gomis:1994rz},
\be \label{constraintdiffs}
\partial_2 F_1 = \partial_1^2 F_1 - \frac{2}{3}\partial_1 F_1 \left(\frac{\partial_1 J}{J} +4\mathcal{L}\partial_1 F_2\right) + 8\partial_1^2 F_2 \mathcal{W},\quad \partial_2 F_2 = \partial_1^2 F_2 - \frac{2}{3}\partial_1 F_2 \frac{\partial_1 J}{J} + (\partial_1 F_1)^2 + \frac{4}{3}(\partial_1 F_2)^2 \mathcal{L},
\ee
where we focussed on the case in which we do a diffeomorphism of the configuration with $\mathcal{L}$ and $\mathcal{W}$ constant. This is appropriate for the geometries we are interested in. Here $J$ is the determinant of the Jacobian of the transformation and given by 
\be 
J = \partial_1 F_1 \partial_2 F_2 - \partial_1 F_2 \partial_2 F_1\, .
\ee
Let us make a few comments. First, even though \eqref{constraintdiffs} depends on $J$, this combination has no explicit $J$ dependence, as it should. Second, the fact that the constraint \eqref{constraintdiffs} depends on $\mathcal{L}$ and $\mathcal{W}$ makes the reduction to a single coordinate $t_1$ non-linear. 

After a diffeomorphism, the Christoffel symbols $\G^{\mu}_{\nu\rho}$ change \footnote{For a diffeomorphism sending $t_\mu(\tilde{t}_1,\dots,\tilde{t}_{N-1})$ this change is given by
\be
\tilde{\G}^{\a}_{\b\g} = \frac{\partial \tilde{t}^{\a}}{\partial t^\rho} \left( \frac{\partial t^{\mu}}{\partial \tilde{t}^{\beta}}\frac{\partial t^{\nu}}{\partial \tilde{t}^{\g}} \G^{\rho}_{\mu\nu} + \frac{\partial^2 t^{\rho}}{\partial \tilde{t}^{\b} \partial \tilde{t}^\g} \right),
\ee
with $\frac{\partial \tilde{t}^{\a}}{\partial t^\rho}$ the inverse of the Jacobian of the diffeomorphism.}
and by using the expression for the stress tensor in \eqref{LGamma}, the transformed stress tensor reads, 
\begin{align}\label{finiteTSchw}
\mathcal{L}[F_1,F_2] = \frac{1}{4(\partial_1 F_1 \partial_2 F_2 - \partial_1 F_2 \partial_2 F_1)}\left(\partial_1 F_1 \partial_2^2 F_2 - \partial_2^2 F_1 \partial_1 F_2 - 2(\partial_1\partial_2 F_1 \partial_2 F_2 - \partial_2 F_1 \partial_1 \partial_2 F_2)\right.\nonumber\\
\left. + 3 \partial_1 F_1 (\partial_2 F_1)^2 + 4(\partial_1 F_1 (\partial_2 F_2)^2 + 2 \partial_2 F_1 \partial_1 F_2 \partial_2 F_2)\mathcal{L}-24 \partial_1 F_1 (\partial_2 F_2)^2 \mathcal{W}\right).
\end{align}
This can be put in a form that purely has $t_1=u$ derivatives by using \eqref{constraintdiffs}. In that case it is the non-linear version of the action we considered in section \ref{sec:disktrumpet} or in other words the generalization of the usual Schwarzian derivative. The boundary action is 
\be \label{fullaction}
S_\partial[F_1,F_2] = -8\g \int_0^\b \d u \,\mathcal{L}[F_1,F_2].
\ee
Here, again, we have in mind having replaced all the $t_2$ derivatives using \eqref{constraintdiffs}. To understand the fields $F_i$ a bit better, let us consider the saddles and fluctutations of this action and show that they are equivalent to the ones obtained in \eqref{sec:disktrumpet}. Let us start by reproducing the $N=2$ theory from this. This is simply the sector with $F_2 = 0$ and $F_1(t_1,t_2) = f(u)$ (after imposing \eqref{constraintdiffs}. We have 
\be 
\mathcal{L}[f] = \mathcal{L} f'^2 - \frac{1}{2} \{f,u\}
\ee
which has the conventional saddle $f = u$ and $\mathcal{L} = \pi^2/\b^2$ in case of the disk. Here $f$ has the boundary condition $f(u+\beta) \sim f(u) + \beta$ and suggests that $F_1 $ is a compact field. The range of $F_2$ can be understood by considering the saddles and fluctutations of the full action. The saddles are given by $F_1 = u$ and $F_2 = t_2$, which suggests that $F_2$ inherits the periodicity of the base space $t_2$ direction. We will come to this below. To find the quadratic action for the fluctuations, we need to parametrize them in such a way that one becomes an infinitesmial diffeomorphisms in $u$ and the other an infinitesimal spin$-3$ transformation. Expanding the modes $F_i$ around the saddles as 
\be \label{expansion}
F_1(u,t_2) = u + \e(u,t_2) - \partial_u \zeta(u,t_2),\quad F_2(u,t_2) = t_2 + 2 \zeta(u,t_2)
\ee
with $\e$ and $\zeta$ small and performing this expansion in \eqref{fullaction} (after imposing the constraints \eqref{constraintdiffs}) we find that the the on-shell value, the variation of the Lagrangian and the quadratic action all agree with the results from section \ref{sec:disktrumpet}!

\subsection{Integration space} 

One of the interesting features of the Schwarzian theory is that we can go to finite temperature by a simple diffeomorphism that maps the circle to the real line \cite{Maldacena:2016upp},
\be \label{tanmap}
t(u) = \tan \frac{\pi \tau(u)}{\b}.
\ee
Geometrically this is an appealing picture, and we would now want to show that one can do something similar in the $\PSL(3,\mathbb{R})$ case. It will also help us to see what the integration space of the diffeomorphisms $F_i$ is. 

The zero temperature $\PSL(3,\mathbb{R})$ theory is given by setting $\mathcal{L}$ and $\mathcal{W}$ to zero and decompactify $t_1$. It will also prove useful to redefine $F_i$ as $F_1 = f_1$ and $F_2 = f_2 - 1/2 f_1^2$, in which case the zero temperature $\PSL(3,\mathbb{R})$ Schwarzian becomes,
\be \label{zeroTSchw}
\mathcal{L}[f_1,f_2] = \frac{\partial_2^2 f_2 \partial_1 f_1 - \partial_2^2 f_1 \partial_1 f_2 - 2 \left(\partial_2 f_2 \partial_1 \partial_2 f_1 - \partial_2 f_1 \partial_1 \partial_2 f_2\right)}{4(\partial_1 f_1 \partial_2 f_2 - \partial_1 f_2 \partial_2 f_1)}\, ,
\ee
where, again, the $f_i$ are constraint to satisfy \eqref{constraintdiffs}. Inserting this constraint directly one can show that this action enjoys a $\PSL(3,\mathbb{R})$ symmetry,
\be \label{PSL3symm}
f_1 \to \frac{a_{11} f_1 + a_{12} f_2 + a_{13}}{a_{31} f_1 + a_{32} f_2 + a_{33}},\qquad f_2 \to \frac{a_{21} f_1 + a_{22} f_2 + a_{23}}{a_{31} f_1 + a_{32} f_2 + a_{33}}.
\ee
with $a_{ij}$ entries of an $\PSL(3,\mathbb{R})$ matrix. This symmetry is coming from the bulk $\PSL(3,\mathbb{R})$ isometry group. From \eqref{zeroTSchw} one can also obtain a more familiar expression \cite{MARSHAKOV199079} by defining $e_1(t_1) = f_1'(t_1)/f_2'(t_1)$ and $e_2(t_1) = f_2(t_1)$,
\be
\mathcal{L}[e_1,e_2] = \frac{1}{4} \left(\frac{e_1'''}{e_1'} + \frac{e_2'''}{e_2'} - \frac{4}{3} \left(\frac{e_1''}{e_1'}\right)^2 - \frac{4}{3}\left(\frac{e_1''}{e_1'}\right)^2 - \frac{1}{3} \frac{e_1'' e_2''}{e_1'e_2'} \right).
\ee
from which one can also obtain the usual Schwarzian by setting $e_1 = e_2$. 

To go to finite temperature, where we have $\mathcal{L}$ nonzero, we need the analogue of \eqref{tanmap}. As it turns out, and we further elaborate on in appendix \ref{app:Map}, this map is given by 
\begin{align}\label{diffeos1}
f_1(t_1,t_2) &= \sin\left(\frac{2\pi F_1(t_1,t_2)}{\b}\right)\exp\left(- \frac{4\pi^2 F_2(t_1,t_2)}{\b^2}\right),\\\label{diffeos2}
f_2(t_1,t_2) &= 1-\cos\left(\frac{2\pi F_1(t_1,t_2)}{\b}\right)\exp\left(- \frac{4\pi^2 F_2(t_1,t_2)}{\b^2}\right).
\end{align}
Plugging these diffeomorphisms in \eqref{zeroTSchw}, we precisely obtain \eqref{finiteTSchw} with $\mathcal{L} = \pi^2/\b^2$ and $\mathcal{W} = 0$. From this map we also see that this action has an $\PSL(3,\mathbb{R})$ symmetry by composing with \eqref{PSL3symm}. This is completely analogous to the $N=2$ case. For the trumpet case, one can do a similar diffeomorphism to get \eqref{finiteTSchw}, but now with $\mathcal{W}$ non-zero as well, see Appendix \ref{app:Map}.

The maps \eqref{diffeos1} and \eqref{diffeos2} are important to understand the integration space of the diffeomorphisms $F_i$. $F_1$ is compact with period $\b$, whereas $F_2$ appears to be non-compact and valued in $\mathbb{R}$. Thus $F_i$ span a plane, just like the $f_i$. The $F_i$ are polar coordinates, whereas the $f_i$ are planar coordinates. However, in the zero temperature case where we have the coordinates $f_i$, we do not want include any points at infinity. In the finite temperature case we do and we should think of them as coordinates on $\mathbb{RP}^2$ (which is what one gets when taking the plane and including lines at infinity). We thus have,
\be 
V_{T=0} = \frac{\tilde{\textsf{Diff}}(\mathbb{R}^2)}{\PSL(3,\mathbb{R})}
\ee
and for the $F_i$ in case of the disk,
\be 
V_{\rm Disk} = \frac{\tilde{\textsf{Diff}}(\mathbb{RP}^2)}{\PSL(3,\mathbb{R})}.
\ee
The tilde indicates the constraints \eqref{constraintdiffs}. The generalization to general $N$ is straightforward,
\be 
V_{\rm Disk} = \frac{\tilde{\textsf{Diff}}(\mathbb{RP}^{N-1})}{\PSL(N,\mathbb{R})}\qquad . 
\ee
For the trumpet we need to be a bit more careful about the integration space. Below we will give some arguments that it should be 
\be 
V_{\rm Trumpet} = \frac{\tilde{\textsf{Diff}}(\mathbb{T}^{N-1})}{U(1)^{N-1}}.
\ee

\subsection{Geometric interpretation}

In the preceding subsections we saw how one can naturally introduce additional boundary coordinates. These coordinates are coordinates on $\mathbb{RP}^{N-1}$, but it is not what generates translations in this geometry. For $t_1 = u$ we know this, it is generated by the Hamiltonian $Q_2 = \int_0^{\b} \d u \mathcal{L}$, i.e. we have $\iota_V \Omega = \d Q$ for $V$ corresponding to the diffeomorphism $(\delta \e,\delta \zeta) = (\e',\zeta')$. One can wonder whether something similar holds for the other conserved charges\footnote{We thank Andreas Blommaert for bringing this up.}. This is indeed the case. To do this, we first need an expression for $\mathcal{W}[F_1,F_2]$, which can be derived in a similar fashion as how one derives \eqref{finiteTSchw} but now starting with \eqref{WGamma} instead of \eqref{LGamma}. This results in
\be 
\mathcal{W}[F_1,F_2] = \frac{1}{8}\left[ \frac{2}{3} \partial_1 \mathcal{L}[F_1,F_2] - \frac{1}{J}\left( \partial_2 F_1 \partial_2^2 F_2 - \partial_2^2 F_1 \partial_2 F_2 + 4 \mathcal{L} \partial_2 F_1 (\partial_2 F_2)^2 - 8 \mathcal{W} (\partial_2 F_2)^3 \right)  \right]
\ee
Let us consider expanding around the saddles as in \eqref{expansion} to quadratic order. This gives us some the same expressions as when we would have done this starting from the gauge theory description in \ref{sec:setup} and \ref{sec:disktrumpet}. The quadratic part of the spin$-3$ charge, $\int_0^{\b} \d u\mathcal{W}[F_1,F_2]$, reads (after imposing \eqref{constraintdiffs}),
\be \label{Wquadratic}
\int_0^{\b} \d u \mathcal{W}_{(2)}[F_1,F_2] = \frac{1}{24}\int_0^{\b} \d u \left( 72 \mathcal{W} \e'^2 + 128 \mathcal{L}^2 \e' \zeta' - 384 \mathcal{L}\mathcal{W} \zeta'^2 - 40 \mathcal{L} \e''\zeta'' + 24 \mathcal{W}\zeta''^2 + 2 \e'''\zeta''' \right)
\ee
One can now easily check that by using $V$ to be a vector field such that 
\be \label{vectorfieldW}
\begin{pmatrix}
\delta \e\\
\delta \zeta 
\end{pmatrix} = \begin{pmatrix}
\partial_{t_2} \e\\
\partial_{t_2} \zeta 
\end{pmatrix} = 
\begin{pmatrix}
-\frac{1}{3}(16\mathcal{L} \zeta' + \zeta''')\\
\e'
\end{pmatrix}
\ee
we indeed satisfy the relation $\iota_V \Omega = \d Q_3$ with $Q_3$ as in \eqref{Wquadratic} and $\Omega$ given in \eqref{symplecticMeasure_ezeta}. This means that the charge $Q_3$ generates $t_2$ translations, just as $Q_2$ generated $t_1 = u$ translations. We think that this conclusion holds for general $N$, so $Q_k$ generates translations in the $t_{k-1}$ direction for $k = 2,\dots,N$. Notice also that the translations in $t_2$ become non-linear and dependent on the particular saddle one expands around when converting back to the boundary time $u$ as can be seen from the second equality in \eqref{vectorfieldW}. 

Circling back to the matrix model dual we proposed in section \ref{sec:matrixmodel}, the statement that we can consider objects like \eqref{generalizedGibbs} is now purely geometric. Each charge in there is a conserved charge that generates translations in the additional directions and $\mu_k$ can be thought of as the range of that direction. This begs for a bulk interpretation. 

The bulk interpretation for the disk is in fact rather clear, it is just the usual disk geometry and since the spin$-3$ field vanishes, there are no real interpretational issues. These issues do arise for the trumpet, since in that case we will have non-zero spin$-3$ fields in the bulk. We will now show that using the additional times we introduced earlier, one can obtain a geometric picture for the trumpet as well, but it will be one in three bulk dimensions (in the case of $N=3$). This geometry will geometrize the two commuting $U(1)$ symmetries we found from the gauge theory perspective. 

We build this geometry by going back to \eqref{gconn} and insert the extended $\Psi(t_1,t_2)$ solutions for the trumpet, see \eqref{Trumpet1}-\eqref{Trumpet3}. Let us denote this group element by $g_{\rm Trumpet}$. This group element gives a flat connection in terms of the coordinates $t_1$ and $r$, but we can actually extend this by a $t_2$ component, while perserving the flatness in all the three coordinates,
\be 
\tilde{A} = A_{\rm Trumpet} + a e^{-r L_0} A_{t_2} e^{ r L_0} \d t_2 = L_0 \d r + e^{-r L_0} A_{t_1} e^{ r L_0} \d t_1 + e^{-r L_0} A_{t_2} e^{ r L_0} \d t_2
\ee
with $A_{t_1} = A_u$ is given by the usual gauge field for the trumpet, \eqref{AAdS2bc} with $\mathcal{L} = - \frac{1}{16\b^2}(\ell_1^2/4 + \ell_2^2/3)$ and $\mathcal{W} = (4/9 \ell_2^2 - \ell_1^2 \ell_2)/(384\b^3)$ and $A_{t_2}$ is given by \footnote{We could also have introduced a relative constant between $A_{\rm Trumpet}$ and the $t_2$ component of $\tilde{A}$, because any such gauge field is a correct extension of the $2d$ gauge field $A_{\rm Trumpet}$. It parametrizes a normalization of the one-form $\d t_2$.}
\be 
A_{t_2} = g_{\rm Trumpet}^{-1} \partial_{t_2} g_{\rm Trumpet}.
\ee
The reason why this extension makes sense is the following. From the discussion in section \ref{sec:setup} we know what gauge transformations leave the trumpet gauge field $A_{\rm Trumpet}$ invariant. These were the gauge transformations \eqref{modes1}-\eqref{modes3} with $\zeta$ and $\e$ set to a constant. For $\zeta = 0$ one finds that this gauge transformation is the $t_1 = u$ translation symmetry, which one can show from the fact that we can write
\be 
A_{\rm Trumpet} = L_0 \d r + \eta_1 \d t_1
\ee
with $\eta_1$ the gauge transformation and so it is clear that $\eta_1$ leaves $A_{\rm trumpet}$ invariant and furthermore, 
\be 
e_{\rm Trumpet} = \frac{1}{2}(A + A^{\dagger}) = L_0 \d r + \eta_1^+ \d t_1
\ee
where we can write $\eta_1^+ = e_{\mu}\xi^\mu_1$ with $\xi_1 = \partial_u$, which was indeed the requirement for interpretating a gauge transformation as a diffeomorphism. As mentioned before, if one tries to do the same with the the gauge transformation for $\e = 0$, this procedure does not work. Using the additional coordinate $t_2$ we can make this work in fact, because the additional $t_2$ component of the gauge field is precisely the gauge transformation for $\e = 0$, i.e. we can write $\tilde{A}$ as
\be \label{Atilde}
\tilde{A} = L_0 \d r + \eta_1 \d t_1 + \eta_2 \d t_2
\ee
and indeed $\eta_2^+ = 1/2(\eta_2 + \eta_2^{\dagger})$ should be interpreted as the $t_2$ component of the (now) dreibein (using an index $A$ for the three coordinates),
\be
\eta_2 = e_A \xi_2^A 
\ee
with $\xi_2 = \partial_{t_2}$. This means that the symmetries of the trumpet have become geometric (and linear) in the three dimensional bulk parametrized by $r$, $t_1$ and $t_2$. The expression \eqref{Atilde} also makes all the symmetries of $\tilde{A}$ explicit (note that $[\eta_1,\eta_2] = 0$). 

\begin{figure}
    \centering
    \includegraphics[scale=0.4]{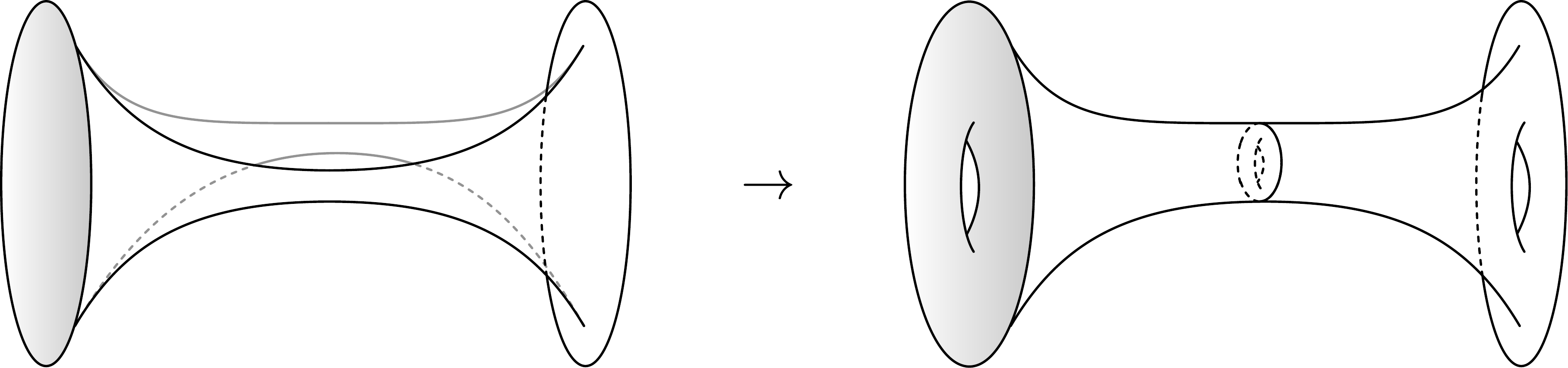}
    \caption{Uplift of the partially non-geometric wormhole contribution (left) in the $2d$ theory to a fully geometric one in $3d$ (right), where the additional higher spin symmetry becomes an ordinary translation symmetry in the $t_2$ coordinate. This gives rise to a $3d$ geometry with the topology of an interval times a torus.}
    \label{fig:DoubleWormhole}
\end{figure}

Now that we understand the gauge field and its symmetries geometrically in three dimensions, we can study what type of bulk metric this gives rise to. Employing the formula, \eqref{metricSpin3}, we find
\be\label{metric3dWormhole}
\d s^2 = \d r^2 + F(r) \d t_1^2 + H(r) \d t_1 \d t_2 + G(r) \d t_2^2, 
\ee
where we have shifted $r$ and rescaled $t_1$ and $t_2$ to make the metric $\b$ independent (this is analogous to what one would do in the $N=2$ case). 
The form of $F, H$ and $G$ is not important. The only thing we need to know is that $F$ and $G$ are positive and $H$ is negative and that the determinant of the metric never degenerates as a function of $r$. This metric thus has two asymptotic boundaries at $r = \pm \infty$ and we want to consider compactifying both $t_1$ and $t_2$ with some range $b_1$ and $b_2$. The resulting $3d$ geometry is our proposal for the geometric wormhole in the $\PSL(3,\mathbb{R})$ theory, see also Fig. \ref{fig:DoubleWormhole}

Let us mention a few facts about this geometry. The metric does not have a constant Ricci curvature, but is negative and only asymptotically constant where it approaches $R = -14$ close to both boundaries. The induced metric on constant $r$ slices is flat, i.e. $R_2 = 0$, whereas the extrinsic curvature of these slices goes smoothly from $-3$ to $3$ (in some units that define the scale of the geometry) when $r$ goes from the left ($r=-\infty$) to right boundary ($r = +\infty$). This means that the extrinsic curvature vanishes somewhere and seems like the natural place to terminate the three dimensional lift of the $\PSL(3,\mathbb{R})$ trumpet. Close to the boundary the metric looks like that of a Lifshitz geometry, because the additional time coordinate $t_2$ scales differently than $t_1$. It would be intersting to see whether one can find a natural candidate for the $3d$ theory that gives rise to these types of geometries. 

\subsection{Generalized Dehn twists}

Using the $3d$ wormhole geometry we can also find more evidence for the generalized Dehn twists we proposed in section \ref{sec:gluing}. We can simply follow the logic in \cite{Saad:2019lba} but we will be more heuristic. To get the measure for the twists and $\ell_i$ we need to insert the coordinate change that implements the twist in the gauge field and compute the symplectic measure. To do so it is actually easiest to consider the gauge field $A$ on the cylinder close to the cutting surface (lets say it is at $r = r_*$), where we can assume we have done a gauge transformation to bring the gauge field in the form,
\be 
A = \begin{pmatrix}
- \d r + \frac{\d y_1}{2} - \frac{\d y_2}{3} & 0 & 0 \\
0 & - \d r + \frac{2\d y_2}{3} & 0 \\
0 & 0 & - \d r - \frac{\d y_1}{2} - \frac{\d y_2}{3} \\
\end{pmatrix}
\ee
which follows from the fact that $\eta_1$ and $\eta_2$ can be simultaneously diagonalized and we have introduced new coordinates $y_i$, which we want to have periodicity $\ell_i$. This gauge field has thus been written in a form where the holonomy, which was previously a holonomy around the boundary in the original theory with one boundary coordinate, has now been written as a product of two holonomies, one around the cycle in the $y_1$ direction and the other around the $y_2$ direction. Thus in some sense the extra coordinate allowed us to geometrize the Cartan directions of $\PSL(3,\mathbb{R})$, just as is the case for $\PSL(2,\mathbb{R})$. Now we implement the twists in the usual way,
\be \label{generalizedDehn}
\d y_i = \ell_i \d x_i + \tau_i \delta(r-r_*)\d r,\quad x_i \sim x_i + 1.
\ee
To obtain the measure for $\ell_i$ and $\tau_i$ one would then compute the symplectic measure \eqref{canonicalSymplecticMeasure}, but this involves two-dimensional gauge fields and needs to be suitable extended as well, which would be worthwhile to understand. The important point here is however not to derive the measure, because that was already done in \cite{WPN3} using just the holonomies on the two dimensional surface and gave the measure \eqref{genWP}. Here we want to illustrate that the higher-spin generalizations of the Dehn twists whose existence we argued for in section \ref{sec:gluing} actually have a nice geometric origin as ordinary Dehn twists but where we extended the two-dimensional bulk with an additional auxiliary coordinate. 

Our arguments have been slightly heuristic here and were based on a proposal for how to extend the $2d$ theory to $3d$. A few comments are in order. 
\begin{enumerate}
    \item In \eqref{generalizedDehn} we wrote a delta function at $r = r_*$. We interpret $r_*$ as the natural place to cut the geometry, which would be at the place the extrinsic curvature of the constant $r$ slices vanishes, which exists as mentioned above. 
    \item The generalisation to general $N$ is straightforward. We would have a $N$ dimensional bulk geometry with $N-1$ isometries and likewise the geometry has the topology $\mathbb{R} \times \mathbb{T}^{N-1}$ with $N-1$ different Dehn twists. 
    \item Our proposal for the $N-1$ dimensional geometry involves compact $t_i$ directions on the boundary, this means that the integration space for the $\PSL(N,\mathbb{R})$ trumpet is $\tilde{\textsf{Diff}}(\mathbb{T}^{N-1})/U(1)^{N-1}$. It would be worthwhile to show this rigoriously using coadjoint orbits of $\Wc_N$ algebras.
\end{enumerate}

\section{Further developments \& Discussion}\label{sec:discussion}

We have seen that higher spin fields in the bulk of asymptotically AdS$_2$ space give rise to a rather different behaviour for the ramp region in the (generalized) spectral form factor. In the bulk this was the result of a quotient by non-trivial large higher spin diffeomorphims, which were generalizations of the usual Dehn twist. On the boundary we proposed a matrix model of $N-1$ commuting matrices and showed that the fluctuations around a saddle have a two-point function that also gives rise to the $T^{N-1}$ behaviour found in the bulk. 

To deepen the analogy with the usual JT gravity story, we also constructed the $\PSL(3,\mathbb{R})$ Schwarzian theory, which, when expanded around a saddle gives the same on-shell action and fluctuations as found from the BF theory analysis. In this analysis it was important to introduce another time variable in order to linearize the problem and view the $\PSL(3,\mathbb{R})$ Schwarzian theory on similar footing as its $\PSL(2,\mathbb{R})$ cousin. These additional times also allowed us to give a more geometric meaning to the higher spin theories and given more evidence for the existence of higher spin generalizations of the Dehn twists.

There are many things we have not touched upon or left rather open. Here we discuss these points further.

\subsection*{Higher genus \& convex projective structures}

Most of our discussion in the main text revolved around simple topologies, such as the disk and cylinder. Based on those calculations we proposed a matrix model dual to the bulk theory. To further strenghten our proposal, and bring it on similar status as the $N=2$ duality \cite{Saad:2019lba}, we would want to find contributions on other topologies as well. Let us mention a few challanges in that regard from both the bulk $\PSL(N,\mathbb{R})$ BF theory and boundary (matrix model) perspective. 

In contrast to the $N=2$, there is no unique (modulo conjugation) $\PSL(N,\mathbb{R})$ connection on the three holed sphere. As mentioned before, there are $(N-2)(N-1)$ internal parameters and consequently the volume of moduli space $V_{g=0,n=3}$ is not simply one anymore. This means that, although we can still perform a pair-of-pants decomposition of the underlying Riemann surface, the pair-of-pants pieces are more complicated to deal with and in particular have explicit dependence on the internal parameters. In the past there has been much progress into how one can efficiently parametrize this moduli space, starting with the work of Goldman \cite{Goldman1990} and later by Fock and Goncharov \cite{fock2006moduli}, see for instance the reviews  \cite{wienhard2018invitation, burger2011higher} and references therein. 

Actually, let us go into a bit more detail about this. Consider the $N=3$ case and a Riemann surface $\Sigma$ with $n$ holes and genus $g$. When Hitchin wrote down a generalisation of Teichmüller theory, it was realised in \cite{GoldmanChoi} that the Hitchin component $N=3$ is equal to the space convex $\mathbb{RP}^2$ structures on $\Sigma$. A convex $\mathbb{RP}^2$ structure on a surface $\Sigma$ means that we can write $\Sigma = \Omega/\G$, where $\Omega \subset \mathbb{RP}^2$ is a convex domain and $\G \subset \PSL(3,\mathbb{R})$ acting without fixed points on $\Omega$. The quotient $\Omega/\G$ thus doesn't contain any orbifold points. The merit of this point of view is that one can understand the various invariants geometrically. The enlarged set of coordinates (compared to the $N=2$) can then be understood essentially as the statement that there are more projective invariants\footnote{This means that they are invariant under the action of $\PSL(3,\mathbb{R})$ or $\PSL(N,\mathbb{R})$ more generally.}, besides the cross-ratio, one can define, such as triple ratios \cite{fock2006moduli} (see also \cite{Labourie_2009, huang2019mcshane, sun2020volume, bonahon2014parameterizing, sun2020flows} and also \cite{WPN3} for an interpretation of the Fenchel-Nielsen length coordinates in terms of projective invariants). From the usual moduli space \eqref{ModuliSpace} point of view, these describe invariants of the representations $\rho: \pi_1(\Sigma) \to \PSL(3,\mathbb{R})$.

Not only is this perspective useful for understanding the parametrization of moduli space in a more (projective) geometric sense, it also useful for constructing generalisations of the Mirzakhani-McShane identities. For instance in \cite{Labourie_2009, huang2019mcshane} these were written down for all $N$, but explicitly so for $N=3$. In case of the once punctured torus it takes the form \cite{huang2019mcshane},
\be 
\sum_{\g} \frac{1}{1+e^{\ell_1(\g) + \tau(\g)}} = 1
\ee
where the sum is over oriented simple closed curves up to homotopy on the punctured torus, just as in the usual McShane identity for the punctured torus. Here $\ell_1$ the generalized Fenchel-Nielsen length coordinates and $e^{\tau(\g)}$ a triple ratio. One can express $\tau$ in terms of the Fenchel-Nielsen coordinates and the internal coordinates

Using these expressions, one could in principle, following Mirzakhani \cite{mirzakhani2007simple}, calculate the volumes of the $N=3$ moduli space. These volumes will all be infinite, because the twist variables are not fully compactified after quotienting by the mapping class group. With our proposal presented in section \ref{sec:gluing} this could be made finite, but there is still the internal coordinates to worry about. Naively these still appear to be unbounded.

Enlarging the mapping class group seems to be the most natural step towards making the volumes finite, not only mathematically, but also from a physical point of view, because in the higher spin theory we want to quotient by not only large diffeomorphisms, but also their higher spin analogues. It would be interesting to find them and understand their action on the moduli space. See \cite{2018arXiv181211199F, thomas2021higher} for some efforts in this direction by viewing the mapping class group elements as symplectomorphisms (recall that the moduli space is a symplectic manifold and that the twist flows arise from the length functions $\ell_i$ as Hamiltonians). See also \cite{seidel2003lectures} for a concrete discussion about generalized Dehn twists, which is perhaps relevant for making our proposal more rigorous.

\subsection*{Topological recursion}

Our matrix model perspective could also provide a fruitful way for trying to find an analogue of the Mirzakhani recursion for $\PSL(N,\mathbb{R})$ BF theory. In JT these recursion relations arise from the topological recursion relation found by Orantin and Eynard \cite{Eynard:2007kz}. In the case of multi-matrix models such recursion also exist, but usually involve a particular $\Tr M_i M_{i+1}$ type interaction between the matrices and are of course not constraint to commute. Furthermore, often times one focuses on resolvents of one of the matrices in the matrix model as for instance in the case of the minimal string matrix models. In our matrix model we have still an arbitrary interaction between the matrices and so we cannot use these results and moreover the spectral density we are after depends on all charges, not just one and so we cannot focus on correlators of one of the matrices. It is unclear whether or not the standard random matrix theory techniques involving resolvents etc. are going to be helpful here, but it would be very interesting to study this further. It would also be helpful to develop the loop equations for this matrix model. 

A different approach could be one along the lines of \cite{Berenstein:2008eg} and instead of implementing the commuting matrices from the start, view it as the strong coupling limit of a conventional hermitian matrix model deformed by interactions of the form $g^2 \sum_{a<b} \Tr [\mathcal{Q}_a,\mathcal{Q}_b]^2$. This would require extreme control of the matrix model and moreover one would want to go beyond the gaussian regime studied in \cite{Berenstein:2008eg}.  

\subsection*{Large $N$ limit}

The theories we studied always contained a finite number of higher spin fields, carrying spin from $2$ to $N$. It is well-known that such systems sometimes have problems, such as violating the chaos bound \cite{Perlmutter:2016pkf}. Taking $N$ to infinity can cure some of those undesired properties as the theory has an infinite number of conserved charges and could become integrable. We have a few comments.

Here we worked with the gauge group $\PSL(N,\mathbb{R})$ and to reach the $N=\infty$ point, it is perhaps more fruitful to start from the get-go with an infinite number of spins and consider a BF theory with the higher spin algebra $\textsf{hs}[\l]$ as the gauge algebra. The commutation relations in this case are known and so in principle one could follow the same approach as in section \ref{sec:setup}. This will result in an infinite number of modes to be integrated over and correspondingly for each spin $s$ the relevant zero modes need to be projected out. Thus one gets a double infinite product that needs regularization,
\be
Z_{\rm Disk}(\b) \sim e^{\frac{a}{\b}} \prod_{s=2}^{\infty} \prod_{n=s}^{\infty} \frac{\b}{n} \sim \b^{5/12} e^{\frac{a}{\b}}
\ee
with $a$ a positive constant and we used zeta function regularization.\footnote{The power $+5/12$ is also what one finds by studying CFTs with $\Wc_{\infty}[\l]$ symmetry. It follows from the asymptotics of the MacMahon function \cite{Castro:2010ce}.} This disk partition function has a positive power of $\b$ as the one-loop factor, which means the inverse laplace transform is not well-defined. If one would define it by analytic continuation, one would have a non-integrable singularity at small energies in the density of states.

A simple but adhoc way of resolving this is by adding a spin one field, say a $U(1)$ gauge field in the bulk decoupled from the other fields. So we imagine just adding another BF term to the action but now with gauge group $U(1)$ \cite{2020, Iliesiu:2019lfc}. This gives a power $-1/12$ instead of $5/12$ and so the density of states, although it blows up at small energies, will be integrable there. 

For the double trumpet we then also get an infinite number of integrals over the twist and length variables, which again need regularization and we get a cylinder contribution\footnote{Again here we imagine adding a spin one field, but since that double trumpet partition function just depends on $\b_1 + \b_2$, it will not give a time-dependence after analytic continuation.},
\be 
Z_{\rm Cyl}^{N\rm=\infty} \sim \prod_{m\geq 1} \frac{\sqrt{\b_1\b_2}}{\b_1 + \b_2} = \frac{\sqrt{\b_1+\b_2}}{(\b_1\b_2)^{1/4}} \to \frac{\sqrt{\b}}{T^{1/2}}.
\ee

For the spectral form factor this has the following implications. There would be an initial exponential decay from the disconnected part, which after an order one amount of time turns into a growth because of the one-loop factor. This is actually consistent with an integrable theory.\footnote{We thank G\'{a}bor S\'{a}rosi for making this suggestion.} The cylinder contribution therefore seems to never dominate, but perhaps the sum over all higher genus corrections causes the spectral form factor to oscillate at late times and approach the plateau. 
    
This signature of integrability could also emerge from our matrix model dual, because in the large $N$ limit, we get an infinite number of matrices and so it becomes a matrix quantum mechanics. This theory is gauged in order to enforce the commutator between any two matrices to vanish. Specifically it would enforce $H$ and $\dot{H}$ to commute. Furthermore, depending on the interaction between the different matrices present in the potential, one could obtain different kinetic terms. The canonical kinetic term arises from the interaction $\Tr H_{a} H_{a+1}$. The large $N$ limit also obscures a bit what we mean with the boundary theory, because there will be an additional boundary coordinate now (unrelated to the one presented in \ref{sec:PSL3Schw}). The bulk then seems to be three dimensions? It would be worthwhile to understand this limit better and its relation to integrability. 

\subsection*{Relation with SYK?}

Another interesting direction that we have not explored here is what the relation with the SYK model is. Are there SYK models that have the Schwarzians proposed in section \ref{sec:PSL3Schw} in the IR? In the usual SYK models we have a bulk that contains massive fields with masses that are order one and to make contact with our theory, some of these masses need to vanish. One can also turn this around and see whether in the BF theory, we can gap some of the higher spin fields. For instance we saw that setting $\partial_1 F_2 = 0$ in the $\PSL(3,\mathbb{R})$ theory we got the usual finite temperature Schwarzian theory, so that seems analogous to gapping the spin$-3$ boundary mode. 

Another point to make is that recently it has been found that in the integrable SYK model, the $q=2$ model, there is an exponential ramp \cite{Winer:2020mdc}, i.e. instead of a linear-in-time behaviour one has $e^T$ due to an additional frequency space symmetry that emerges when $q=2$. In our model we can also see an exponential ramp. The dip and Heisenberg time are 
\be 
T_{\rm dip} \sim e^{\frac{2\S}{(N-1)(N+2)}} ,\quad T_{\rm Heis} \sim e^{\frac{2\S}{N-1}}
\ee
So the region in time that the cylinder dominates becomes smaller and smaller as $N$ increases, but still the spectral form factor needs to reach a value of order $e^{\S}$ (height of the plateau) and so it needs to increase rapidly. To access this one could take the general $N$ answer \eqref{SFFHS} and scale $T/\b$ as $T/\b = a e^{\tau/(N-1)} = a\left(1 + \frac{\tau}{N-1}\right)$ with $a$ an order one constant and $\tau$ a dimensionless parameter that does not scale with $N$. This results in
\be 
c_N \left(1 + \frac{T^2}{\b^2}\right)^{\frac{N-1}{2}} \Rightarrow c_N \left(1 + a^2 + \frac{2 a^2 \tau}{N-1}\right)^{\frac{N-1}{2}} \sim \tilde{c}_{N,a} e^{\frac{a^2}{1+a^2}\tau}
\ee
Thus we get an exponential behaviour in the dimensionless time $\tau$. It would be worthwhile to deepen the relation with $q=2$ SYK and large $N$ higher spin theories more, if there is any. Notice that this is a different limit than the one discussed above where we start with the $\textsf{hs}[\lambda]$ theory directly, because there we were in the strict $N=\infty$ limit.

Finally, related to the comment earlier about the large $N$ limit giving rise to another dimension, one can wonder whether the large $N$ limit is in fact related to a two dimensionsal version of the SYK model \cite{Turiaci:2017zwd}.

\subsection*{Lorentzian musings}

\begin{figure}
    \centering
    \includegraphics[scale=0.4]{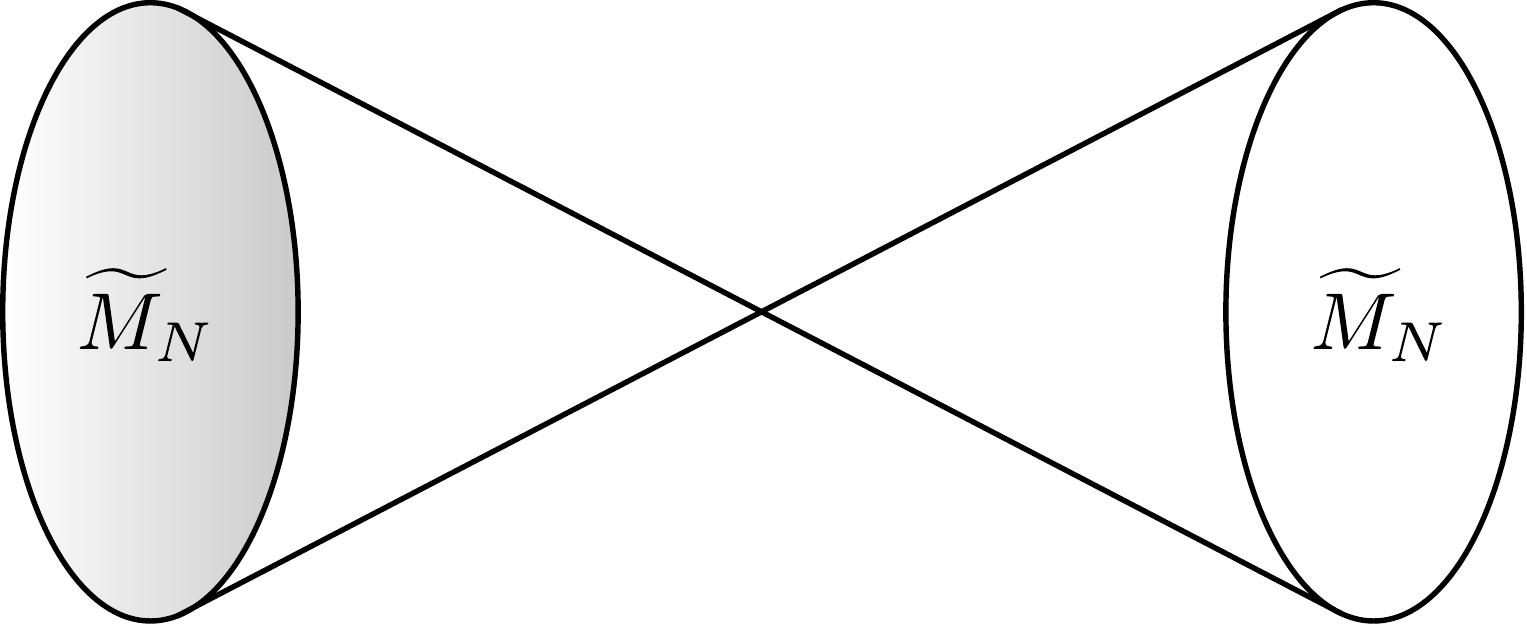}
    \caption{Possible generalization of the double cone geometry to the $\PSL(N,\mathbb{R})$ theory. The boundary geometry is the higher dimensional $M_N$ manifold introduced in section \ref{sec:PSL3Schw}, but suitably analytically continued (signified by the tilde). }
    \label{fig:DoubleCone}
\end{figure}

The calculations done in the main text were done purely in Euclidean signature and continued to Lorentzian signature afterwards. Just as was done JT gravity \cite{Saad:2018bqo}, one can also wonder about the Lorentzian geometry that gives the $T^{N-1}$ behaviour. Recall that the linear in $T$ behaviour for JT arose from a zero mode in the double cone geometry originated from a relative origin of time on both sides. To get the $T^{N-1}$ behaviour then suggests that one should be looking for more zero modes. The natural place to look for them is in the additional boundary coordinates introduced in section \ref{sec:PSL3Schw}. After a suitable quotient they would give a higher-dimensional volume for the zero modes on the boundary and perhaps this could give rise to the higher powers of $T$, see Fig. \ref{fig:DoubleCone}.

\section*{Acknowledgements}

I am happy to thank Onkar Parrikar for initial collaboration and Andreas Blommaert, Jan de Boer, Shouvik Datta, Luca Iliesiu, Raghu Mahajan, Edward Mazenc, G\'{a}bor S\'{a}rosi, Steve Shenker, Douglas Stanford, Alexander Thomas, Joaquin Turiaci and Zhenbin Yang for discussions. We especially grateful to Lotte Hollands, Yi Huang, Fran\c{c}ois Labourie and Zhe Sun for discussion and correspondence on the higher Fenchel-Nielsen coordinates, convex $\mathbb{RP}^2$ structures and their volumes. I am is supported by the Simons Foundation. 

\appendix

\section{$\PSL(3,\mathbb{R})$ generators}\label{app:generators}
The generators of $\PSL(3,\mathbb{R})$ we are using are,
\begin{align}
L_1 &= \begin{pmatrix}
0 & 0 & 0 \\
\sqrt{2} & 0 & 0\\
0 & \sqrt{2} & 0
\end{pmatrix},\quad L_0 = \begin{pmatrix}
1 & 0 & 0 \\
0 & 0 & 0\\
0 & 0 & -1
\end{pmatrix},\quad 
L_{-1} = \begin{pmatrix}
0 & -\sqrt{2} & 0 \\
0 & 0 & -\sqrt{2}\\
0 & 0 & 0
\end{pmatrix}\nonumber\\
W_2 &= \begin{pmatrix}
0 & 0 & 0 \\
0 & 0 & 0\\
4 & 0 & 0
\end{pmatrix},\quad W_1 = \begin{pmatrix}
0 & 0 & 0 \\
\sqrt{2} & 0 & 0\\
0 & -\sqrt{2} & 0
\end{pmatrix},\quad 
W_0 = \frac{2}{3}\begin{pmatrix}
1 & 0 & 0 \\
0 & -2 & 0\\
0 & 0 & 1
\end{pmatrix}\nonumber\\
W_{-1} &= \begin{pmatrix}
0 & -\sqrt{2} & 0 \\
0 & 0 & \sqrt{2}\\
0 & 0 & 0
\end{pmatrix},\quad W_{-2} = \begin{pmatrix}
0 & 0 & 4 \\
0 & 0 & 0\\
0 & 0 & 0
\end{pmatrix}
\end{align}
For general $N$ we also note \cite{Gonzalez:2018enk}
\be 
L_1 = \begin{pmatrix}
0 & 0 & 0 & \dots & 0 \\
\sqrt{k_1} & 0 & 0 & \dots & 0\\
0 & \sqrt{k_2} & 0 &  \dots & 0 \\
\vdots & \ddots & \ddots & \ddots & \vdots \\
0 & \dots & 0 & \sqrt{k_{N-1}} & 0 
\end{pmatrix},\quad 
L_{-1} = \begin{pmatrix}
0 & -\sqrt{k_1} & 0 & \dots & \dots & 0 \\
0 & 0 & -\sqrt{k_2} & 0 & \dots & 0\\
\vdots & \ddots & \ddots & \ddots & \ddots & \vdots \\
0 & \dots & \dots & \dots & 0 & -\sqrt{k_{N-1}}\\
0 & \dots & \dots & \dots & \dots & 0
\end{pmatrix}\, , 
\ee
with $k_i = 2 \sum_{j} (K^{-1})_{ij}$ and $K_{ij}$ the Cartan matrix of $\textsf{SL}(N,\mathbb{R})$, 
\be 
K_{ij} = \left\{ \begin{array}{ll}
     2 & i = j \\
     -1 & |i - j| = 1\\
     0 & {\rm otherwise}
\end{array}\right. . 
\ee

\section{Going from zero to finite temperature}\label{app:Map}

In this appendix we explain where the maps \eqref{diffeos1} and \eqref{diffeos2} come from. It will be helpful to first review how one can derive \eqref{tanmap} in the $N=2$ case. In this case there are no additional boundary coordinates and we have the following equation for the $\Psi_i$ (focussing on constant $\mathcal{L})$,
\be 
\partial_u^2 \Psi_i(u) + \mathcal{L}\Psi_i(u) = 0,
\ee
which has two linear independent solutions $\Psi_1 = \sin(\pi u/\b)$ and $\Psi_2 = \cos(\pi u/\b)$ that are consistent with the holonomy being $-\mathbb{1}$. The coordinate $\Psi_1/\Psi_2$ is a homogeneous coordinate on $\mathbb{RP}^1$. This coordinate takes real values. By considering a diffeomorphism $\tau(u)$ of the boundary circle, we can write
\be 
\frac{\Psi_1(u)}{\Psi_2(u)} = \tan \frac{\pi \tau(u)}{\b}
\ee
Denoting the LHS by $t(u)$ we get the map \eqref{tanmap} sending the circle to the real line. In fact, the circle is topologically equivalent to the real projective line, so we get a map from $\mathbb{RP}^1$ to $\mathbb{R}$. In other words for the integration space of the finite temperature Schwarzian theory we can equivalently write $\textsf{Diff}(\mathbb{RP}^1)/\PSL(2,\mathbb{R})$ (and $\textsf{Diff}(\mathbb{R})/\PSL(2,\mathbb{R})$ at zero temperature), which is the more natural object to generalize to arbitary $N$ as mentioned in the main text. 

For the $N=3$ case we want to proceed in exactly the same way. Let us focus on the disk geometry, in which case we want to solve \eqref{reduction} with $\mathcal{L} = \pi^2/\b^2$ and $\mathcal{W} = 0$,
\be \label{Diskflow}
\partial_{t_1}^3 \Psi_i  + \frac{4\pi^2}{\b^2} \partial_{t_1} \Psi_i = 0,\quad \partial_{t_2} \Psi_i = \partial_{t_1}^2 \Psi_i + \frac{8\pi^2}{3\b^2} \Psi_i
\ee
The linear independent solutions consistent with the trivial holonomy are 
\begin{align}
\Psi_1(t_1,t_2) &= \frac{\beta}{\sqrt{2}\pi}\sin\left(\frac{2\pi t_1}{\b}\right)e^{- \frac{4\pi^2 t_2}{3\b^2}},\\
\Psi_2(t_1,t_2) &= \frac{\beta}{\sqrt{2}\pi}e^{\frac{8\pi^2 t_2}{3\b^2}}\left(1 - \cos\left(\frac{2\pi t_1}{\b}\right)e^{- \frac{4\pi^2 t_2}{\b^2}}\right),\\
\Psi_3(t_1,t_2) &= \frac{\beta}{\sqrt{2}\pi}e^{\frac{8\pi^2 t_2}{3\b^2}}.
\end{align}
Here we also chose the $\Psi_1$ and $\Psi_2$ so that at the zero temperature limit solutions they reduce to solutions of \eqref{reduction} with $\mathcal{L} = 0 = \mathcal{W}$. The normalization of each solution is chosen so that the group element constructed from the $\Psi$'s has unit determinant. The ratios $\Psi_i/\Psi_3$ then take the form \eqref{diffeos1} and \eqref{diffeos2} and so we get a map from $\mathbb{RP}^2$ to $\mathbb{R}^2$. 

It is also straightforward to repeat this for the trumpet geometry where we have 
\be 
\mathcal{L} = - \frac{1}{16\b^2} \left(\frac{\ell_1^2}{4} + \frac{\ell_2^2}{3}\right),\quad \mathcal{W} = - \frac{1}{64\b^3} \frac{\ell_2(9\ell_1^2 - 4 \ell_2^2)}{54}
\ee
The linear independent solutions of \eqref{reduction} with the required hyperbolic holonomy and such that the group element $g$ has unit determinant are, 
\begin{align}\label{Trumpet1}
\Psi_1(t_1,t_2) &= \sqrt{8}\b e^{\frac{A_- t_2 + B_- t_1 \b}{12\b^2}},\\\label{Trumpet2}
\Psi_2(t_1,t_2) &= \sqrt{8}\b e^{\frac{A_0 t_2 + B_0 t_1 \b}{12\b^2}},\\\label{Trumpet3}
\Psi_3(t_1,t_2) &= \frac{8\sqrt{2}\b}{\ell_1(\ell_1^2 - 4 \ell_2^2)} e^{\frac{A_+ t_2 + B_+ t_1 \b}{12\b^2}},
\end{align}
with 
\begin{align}
A_{\pm} &= \frac{\ell_1^2}{4} - \frac{\ell_2^2}{3} \pm \ell_1 \ell_2,\quad A_0 = \frac{1}{6}(4\ell_2^2 - 3 \ell_1^2)\\
B_{\pm} &= -(\pm 3\ell_1 + 2 \ell_2),\quad B_0 = 4\ell_2.
\end{align}
We picked the coefficients so that the ratios $\Psi_i/\Psi_3$ go to zero on the boundary of $\mathcal{R}$ in \eqref{spaceR}. 

\section{Grand canonical ensemble}\label{app:GCE}

In the main text we considered a simple boundary action, which just involves the spin-$2$ charge $\mathcal{L}$. This is the standard canonical ensemble where we fix the temperature. Since we also have other conserved charges, we can also consider turning on a chemical potential for those. We do this here. Probably this action with the chemical potential turned on is not one-loop exact, so the calculation done here is just to show how to set up the boundary conditions to get such chemical potentials. 

To turn on these higher chemical potentials, we need to change the boundary condition for the field $B$ and the boundary action in order to ensure a well-defined variational principle. We will focus on Dirichlet boundary conditions, so we look at fixed chemical potentials. Let us consider the boundary condition 
\be \label{grandcanonicalBC}
B = - i c_2 A_u - i c_3 \left( A_u^2 - \frac{\Tr A_u^2}{3} \right).
\ee
which satisfies the equation of motion $[B,A_u] = 0$. Inserting this in the variation of \eqref{action2d}, we get
\begin{align}\label{variationGCE} 
\delta S &= ({\rm EOM}) + \int \Tr \left[ c_3 A_u^2 \delta A_u + \frac{1}{2}\delta c_2 A_u^2 + \frac{1}{2}\delta c_3 A_u^3 \right]\\
&= ({\rm EOM}) + \int \Tr \left[ \frac{1}{3}\delta\left( c_3 A_u^3\right) + \frac{1}{2}\delta c_2 A_u^2 + \frac{1}{6} \delta c_3 A_u^3 \right]
\end{align}
for fixed chemical potentials $c_i$ this does not vanish unless we add the boundary term 
\be 
-\frac{c_3}{3} \int \Tr A_u^3
\ee
to $S$. The total boundary action that makes our variational principle well-defined with boundary condition \eqref{grandcanonicalBC} is thus
\be 
S_b = \frac{i}{2} \int \Tr B A_u - \frac{c_3}{3} \int \Tr A_u^3 = \int \frac{c_2}{2} \Tr A_u^2 + \frac{c_3}{6} \Tr A_u^3
\ee
The $c_i$ are free parameters at this point and define the ensemble, just as fixed temperature defines the canonical ensemble. We can always absorb $c_2$ in a redefinition of $u$ and normalize $c_3$ such that
\be 
S_b = \g \int \Tr [ A_u^2 + \mu A_u^3]
\ee
This boundary action will not only change its on-shell value as compared to the one in the previous subsection, but also the action of the fluctuations. The gauge field $A_u$ still has the same form, i.e. the boundary conditions for $A_u$ did not change and $S$ in \eqref{S2} becomes, 
\be 
S' = S + \g \m \int_0^\b du \left( 72 \mathcal{W} \e'^2 - 384 \mathcal{L}\mathcal{W} \zeta'^2 + 24 \mathcal{W} \zeta''^2 + 128 \mathcal{L}^2 \e' \zeta' - 40 \mathcal{L} \e'' \zeta'' + 2 \e'''\zeta''' \right).
\ee
The measure remains the same, because $A_u$ did not change and so the change in the partition function comes from the change of the gaussian integrals over the fluctuations. In terms of modes, $S'$ is given by
\begin{align}
S' = S &+ 2\g \mu \left( \sum_{n\geq n_0} \frac{288 n^2 \pi^2}{\b} \mathcal{W} ((\e_n^R)^2 + (\e_n^I)^2) + \sum_{n\geq m_0} \frac{384 \pi^4 n^2 \mathcal{W}}{\b^3}\left(n^2 - \frac{4\mathcal{L}\b^2}{\pi^2}\right) ((\zeta_n^R)^2 + (\zeta_n^I)^2)  \right. \nonumber \\
& + \left.\sum_{n\geq m_0} \frac{128 \pi^6 n^2}{\b^5}\left(n^2 - \frac{\mathcal{L}\b^2}{\pi^2}\right)\left(n^2 - \frac{4\mathcal{L}\b^2}{\pi^2}\right) (\e_n^R \zeta_n^R + \e_n^I \zeta_n^I) \right)
\end{align}
The gaussian integrals then give the following partition function, 
\be 
Z(\b,\m) = e^{8\g \b \mathcal{L} - 24 \g \mu \b \Wc} \prod_{n\geq m_0} \frac{\b^2 \a^2}{4\g^2 n^2} \frac{1}{1 - \mu F_n} \prod_{n_0 \leq n < m_0} \frac{\a \b}{2 n \g} \frac{1}{1 + \frac{18 \mathcal{W} \b^2 \m }{n^2 \pi^2 - \mathcal{L} \b^2 }}
\ee
with 
\be 
F_n = \frac{3 \pi^2}{\b^2} \left(n^2 - \frac{4\mathcal{L}\b^2}{\pi^2}\right) \frac{ \mu \frac{54 \mathcal{W}^2}{\pi^6} - 3 \left(n^2 - \frac{\mathcal{L}\b^2}{\pi^2}\right) \frac{\mathcal{W} \b^4}{\pi^4} - \mu \left(n^2 - \frac{\mathcal{L}\b^2}{\pi^2}\right)^2 \left(n^2 - \frac{4\mathcal{L}\b^2}{\pi^2}\right)   }{\left(n^2 - \frac{\mathcal{L}\b^2}{\pi^2}\right)^2 \left(n^2 - \frac{4\mathcal{L}\b^2}{\pi^2}\right) - \frac{27 \mathcal{W}^2 \b^6}{\pi^6}}
\ee
The partition function thus looks a lot more complicated. For the disk things simplify since we have $\mathcal{W} = 0$ and $\mathcal{L} = \pi^2/\b^2$ and we obtain
\be 
Z_{\rm Disk}(\b, \mu) = e^{\frac{8\g \pi^2}{\b}}\frac{\a \b}{4 \g} \prod_{n\geq 3} \frac{\b^2\a^2}{4\g^2 n^2}\frac{1}{1 + \m^2\frac{3\pi^2}{\b^2}(n^2 - 4)}.
\ee
The trumpet is much more complicated, since then $\mathcal{W}$ will be non-zero. Again we emphasize that because the symplectic measure and the one-loop determinant did not `almost' cancel, the result is probably not one-loop exact and might receive additional corrections. It is for sure true that $\iota_V \Omega \neq \d H$ with $V$ the vector field generating `time' translations.

\bibliographystyle{ourbst}
\bibliography{Refs}

\end{document}